\documentclass[iop]{emulateapj}

\slugcomment{Manuscript ver. August 20 \\Received 2015 March 12; accepted 2015 August 16}

\defcitealias{2013PASJ...65..100I}{I13} 

\usepackage{color}
\usepackage{apjfonts}
\usepackage{natbib}
\usepackage[dvipdfmx,colorlinks=true,linkcolor=blue,filecolor=blue,urlcolor=blue,citecolor=blue]{hyperref}
\hypersetup
{
	pdfauthor={T. Izumi et al.},
	pdftitle={Izumi2015a}
}

\shorttitle{ALMA Observations of the Active Nucleus of NGC 7469}
\shortauthors{T. Izumi et al.}

\begin{document}

\title{ALMA Observations of the Submillimeter Dense Molecular Gas Tracers \\
in the Luminous Type-1 Active Nucleus of NGC 7469}


\author{Takuma Izumi\altaffilmark{1}, 
Kotaro Kohno\altaffilmark{1,2}, 
Susanne Aalto\altaffilmark{3}, 
Akihiro Doi\altaffilmark{4}, 
Daniel Espada\altaffilmark{5,6,7}, 
Kambiz Fathi\altaffilmark{8}, 
Nanase Harada\altaffilmark{9}, 
Bunyo Hatsukade\altaffilmark{6}, 
Takashi Hattori\altaffilmark{10,6}, 
Pei-Ying Hsieh\altaffilmark{9,11}, 
Soh Ikarashi\altaffilmark{1,12}, 
Masatoshi Imanishi\altaffilmark{10,6,7}, 
Daisuke Iono\altaffilmark{6,7}, 
Sumio Ishizuki\altaffilmark{6}, 
Melanie Krips\altaffilmark{13}, 
Sergio Mart{\'{i}}n\altaffilmark{13}, 
Satoki Matsushita\altaffilmark{9}, 
David S. Meier\altaffilmark{14}, 
Hiroshi Nagai\altaffilmark{6}, 
Naomasa Nakai\altaffilmark{15}, 
Taku Nakajima\altaffilmark{16}, 
Kouichiro Nakanishi\altaffilmark{5,6,7}, 
Hideko Nomura\altaffilmark{17}, 
Michael W. Regan\altaffilmark{18}, 
Eva Schinnerer\altaffilmark{19}, 
Kartik Sheth\altaffilmark{20}, 
Shuro Takano\altaffilmark{21}, 
Yoichi Tamura\altaffilmark{1}, 
Yuichi Terashima\altaffilmark{22}, 
Tomoka Tosaki\altaffilmark{23}, 
Jean L. Turner\altaffilmark{24}, 
Hideki Umehata\altaffilmark{1}, and 
Tommy Wiklind\altaffilmark{5}
}

\affil{${}^{1}$\ Institute of Astronomy, School of Science, The University of Tokyo, 2-21-1 Osawa, Mitaka, Tokyo 181-0015, Japan; \href{mailto:takumaizumi@ioa.s.u-tokyo.ac.jp}{takumaizumi@ioa.s.u-tokyo.ac.jp}}
\affil{${}^{2}$\ Research Center for the Early Universe, The University of Tokyo, 7-3-1 Hongo, Bunkyo, Tokyo 113-0033} 
\affil{${}^{3}$\ Department of Earth and Space Sciences, Chalmers University of Technology, Onsala Observatory, 439 94 Onsala, Sweden}
\affil{${}^{4}$\ Institute of Space and Astronautical Science, 3-1-1 Yoshinodai, Chuo-ku, Sagamihara 252-5210, Japan}
\affil{${}^{5}$\ Joint ALMA Observatory, Alonso de C{\'{o}}rdova, 3107, Vitacura, Santiago 763-0355, Chile}
\affil{${}^{6}$\ National Astronomical Observatory of Japan, 2-21-1 Osawa, Mitaka, Tokyo 181-8588, Japan}
\affil{${}^{7}$\ SOKENDAI (The Graduate University for Advanced Studies), 2-21-1 Osawa, Mitaka, Tokyo 181-8588, Japan}
\affil{${}^{8}$\ Stockholm Observatory, Department of Astronomy, Stockholm University, AlbaNova Centre, 106 91 Stockholm, Sweden}
\affil{${}^{9}$\ Academia Sinica, Institute of Astronomy \& Astrophysics, P.O. Box 23-141, Taipei 10617, Taiwan} 
\affil{${}^{10}$\ Subaru Telescope, NAOJ, 650 North A'ohoku Place, Hilo, HI 96720, USA} 
\affil{${}^{11}$\ Institute of Astronomy, National Central University, No. 300, Jhongda Road, Jhongli City, Taoyuan County 32001, Taiwan, Republic of China} 
\affil{${}^{12}$\ European Southern Observatory, Karl-Schwarzschild-Str. 2, D-85748 Garching, Germany}
\affil{${}^{13}$\ Institut de Radio Astronomie Millim{\'{e}}trique, 300 rue de la Piscine, Domaine Universitaire, 38406 St. Martin d'H{\`{e}}res, France}
\affil{${}^{14}$\ Department of Physics, New Mexico Institute of Mining and Technology, 801 Leroy Place, Soccoro, NM 87801, USA}
\affil{${}^{15}$\ Department of Physics, Faculty of Pure and Applied Sciences, University of Tsukuba, 1-1-1 Ten-nodai, Tsukuba, Ibaraki 305-8571, Japan}
\affil{${}^{16}$\ Solar-Terrestrial Environment Laboratory, Nagoya University, Furo-cho, Chikusa-ku, Nagoya 464-8601, Japan}
\affil{${}^{17}$\ Department of Earth and Planetary Sciences, Tokyo Institute of Technology, 2-12-1 Ookayama, Meguro, Tokyo 152-8550, Japan}
\affil{${}^{18}$\ Space Telescope Science Institute, 3700 San Marine Drive, Baltimore, MD 21218, USA}
\affil{${}^{19}$\ Max Planck Institute for Astronomy, K{\"{o}}nigstuhl 17, Heidelberg 69117, Germany}
\affil{${}^{20}$\ National Radio Astronomy Observatory, 520 Edgemont Road, Charlottesville, VA 22903, USA}
\affil{${}^{21}$\ Department of Physics, General Education, College of Engineering, Nihon University, Tamuramachi, Koriyama, Fukushima 963-8642, Japan}
\affil{${}^{22}$\ Department of Physics, Ehime University, 2-5 Bunkyo-cho, Matsuyama, Ehime 790-8577, Japan}
\affil{${}^{23}$\ Department of Geoscience, Joetsu University of Education, Yamayashiki, Joetsu, Niigata 943-8512, Japan}
\affil{${}^{24}$\ Department of Physics and Astronomy, UCLA, 430 Portola Plaza, Los Angeles, CA 90095-1547, USA}


\begin{abstract}
We present ALMA Cycle 1 observations of the central kpc region of the luminous type-1 Seyfert galaxy NGC 7469 
with unprecedented high resolution (0.5$''$ $\times$ 0.4$''$ = 165 pc $\times$ 132 pc) at submillimeter wavelengths. 
Utilizing the wide-bandwidth of ALMA, we simultaneously obtained HCN(4-3), HCO$^+$(4-3), CS(7-6), 
and partially CO(3-2) line maps, as well as the 860 $\mu$m continuum. 
The region consists of the central $\sim$ 1$''$ component and the surrounding starburst ring with a radius of $\sim$ 1.5$''$-2.5$''$. 
Several structures connect these components. 
Except for CO(3-2), these dense gas tracers are significantly concentrated towards the central $\sim$ 1$''$, 
suggesting their suitability to probe the nuclear regions of galaxies. 
Their spatial distribution resembles well those of centimeter and mid-infrared continuum emissions, 
but it is anti-correlated with the optical one, indicating the existence of dust obscured star formation. 
The integrated intensity ratios of HCN(4-3)/HCO$^+$(4-3) and HCN(4-3)/CS(7-6) are 
higher at the AGN position than at the starburst ring, which is consistent to our previous findings (submm-HCN enhancement). 
However, the HCN(4-3)/HCO$^+$(4-3) ratio at the AGN position of NGC 7469 (1.11$\pm$0.06) 
is almost half of the corresponding value of the low-luminosity type-1 Seyfert galaxy NGC 1097 (2.0$\pm$0.2), 
despite the more than two orders of magnitude higher X-ray luminosity of NGC 7469. 
But the ratio is comparable to that of the close vicinity of the AGN of NGC 1068 ($\sim$ 1.5). 
Based on these results, we speculate that some other heating mechanisms than X-ray 
(e.g., mechanical heating due to AGN jet) can contribute significantly for shaping the chemical composition in NGC 1097. 
\end{abstract}

\keywords{galaxies: active --- galaxies: individual (NGC 7469) --- galaxies: ISM --- ISM: molecules}

\section{Introduction}\label{sec1} 
Investigation of the distribution, kinematics, and mass of dense molecular material 
in the centers of galaxies plays a key role in studying the evolution of galaxies 
because it is the reservoir of fuel for active galactic nuclei (AGNs), and also the site of massive star formation (starburst = SB). 
On the other hand, these heating sources significantly affect their surrounding gas in radiative and/or mechanical ways. 
For example, intense UV radiation from massive stars forms photodissociation regions (PDRs) around them, 
and X-ray dominated regions (XDRs), which can be larger in volume than PDRs, are formed in the vicinity of AGNs 
(e.g., \citealt{1996ApJ...466..561M,1999RvMP...71..173H,2005A&A...436..397M,2007A&A...461..793M}). 
Cosmic rays from frequent supernovae (SNe) and the injection of mechanical energy due to SNe 
or AGN jet/outflow (mechanical heating) are also important for the chemical layout 
(e.g., \citealt{2011A&A...525A.119M,2012A&A...542A..65K,2015A&A...574A.127K,2015ApJ...799...26M}). 
Thus, the dense molecular gas can provide us critical information on the nature of 
the underlying physical processes accompanying the non-thermal radiation in the centers of galaxies. 

Millimeter/submillimeter spectroscopic observations are indispensable for such a study 
since they can trace dense, cold molecular material with little dust extinction, 
which takes the bulk of gas mass in the nuclear regions of galaxies. 
This would contrast with observations of warm molecular gas such as H$_2$ 1-0 S(1) in the near-IR (e.g., \citealt{2009ApJ...696..448H}). 

Nevertheless, our knowledge on the circumnuclear dense molecular matter has still been limited, especially for AGNs. 
This is because there have been only a small number of high resolution (i.e., $\sim$ 100 pc or higher scale) 
and sensitive observations of dense molecular gas even in nearby AGNs 
(e.g., \citealt{2010A&A...519A...2G,2011ApJ...736...37K,2012MNRAS.424.1963S,2015ApJ...799...26M}), 
although several low resolution surveys of dense gas have been conducted 
with single dish telescopes to address their global properties (e.g., \citealt{2004ApJS..152...63G,2004ApJ...606..271G}). 

This situation is now being improved significantly by the advent of the {\it{Atacama Large Millimeter/submillimeter Array (ALMA)}}. 
Indeed, the spatial distribution, kinematics, and the chemistry have been reported in greater detail for, 
e.g., NGC 1068 (\citealt{2014A&A...567A.125G,2014A&A...570A..28V,2014PASJ...66...75T,2015PASJ...67....8N}) 
, NGC 1097 (\citealt{2013PASJ...65..100I,2013ApJ...770L..27F,2015A&A...573A.116M}), 
NGC 1433 (\citealt{2013A&A...558A.124C}), and NGC 1566 (\citealt{2014A&A...565A..97C}). 
Among these works, \citet{2013PASJ...65..100I} (hereafter \citetalias{2013PASJ...65..100I}) suggested 
that HCN(4-3)/HCO$^+$(4-3) and HCN(4-3)/CS(7-6) integrated intensity ratios seem to be 
higher in AGNs than in SB galaxies ({\it{submm-HCN enhancement}}), 
although the number of the sample galaxies in \citetalias{2013PASJ...65..100I} is few 
and the actual cause of the enhancement is not clear. 

With the above things in mind, we here present the results of our high resolution 
ALMA band 7 observations of submillimeter dense gas tracers towards the central kpc region of NGC 7469. 
Because NGC 7469 hosts a luminous type-1 AGN 
($L_{\rm{2-10keV}}$ = 1.5 $\times$ 10$^{43}$ erg s$^{-1}$; \citealt{2014ApJ...783..106L}), 
we can investigate the effects of AGN luminosity on the surrounding molecular gas 
with little uncertainty by directly comparing the results 
with those obtained in the type-1 low-luminosity 
($L_{\rm{2-10keV}}$ = 6.9 $\times$ 10$^{40}$ erg s$^{-1}$; \citealt{2014ApJ...783..106L}) 
AGN, NGC 1097 (\citetalias{2013PASJ...65..100I}). 

\subsection{The target galaxy NGC 7469}\label{sec1.1}
NGC 7469 is an active galaxy with a large scale (several kpc) 
stellar bar detected at near-IR (NIR) wavelength (\citealt{2000ApJ...529...93K}). 
This galaxy is located at the distance of 70.1 Mpc 
($z$ = 0.0164; \citealt{1990ApJ...354..158M}) thus 1$''$ = 330 pc, 
and hosts a luminous type-1 Seyfert nucleus as evidenced by 
broad Balmer emission lines (FWHM of H$\beta$ = 4369 km s$^{-1}$; \citealt{2014ApJ...795..149P}) 
with time variability (e.g., \citealt{1990ApJ...353..445B,1998ApJ...500..162C}). 
Time variability is also confirmed by UV and X-ray observations 
(e.g., \citealt{2000ApJ...535...58K,2000ApJ...544..734N,2004A&A...413..477P,2005ApJ...634..193S}), 
revealing this nucleus is indeed an AGN. 
NGC 7469 is also classified as a luminous infrared galaxy (LIRG) due to its high IR luminosity 
($L_{\rm {8-1000}}$ $\equiv$ $L_{\rm IR}$ = 10$^{11.7}$ $L_\odot$; \citealt{2003AJ....126.1607S}). 

The type-1 AGN is surrounded by a luminous circumnuclear SB ring 
with a radius of 1.5$''$--2.5$''$ or 495--825 pc (e.g., \citealt{2003AJ....126..143S,2007ApJ...661..149D}, see also Figure \ref{figure1}a), 
which accounts for two thirds of the bolometric luminosity of the galaxy (\citealt{1995ApJ...444..129G}). 
Therefore, this galaxy provides us with a nice opportunity to investigate how AGN and SB 
influence their surrounding molecular gas when observed at high resolution. 

The SB ring is prominent at various wavelengths including centimeter 
(\citealt{1991ApJ...378...65C,1991ApJ...381...79W,2010MNRAS.401.2599O}), 
far-IR (FIR, \citealt{2000ApJ...537..631P}), mid-IR (MIR, \citealt{2003AJ....126..143S,2004ApJ...605..156G}), 
NIR (\citealt{1995ApJ...444..129G}), optical-to-UV (\citealt{1998ApJS..117...25M,2000MNRAS.311..120D,2007ApJ...661..149D}), 
and soft X-ray (\citealt{1996ApJ...468..191P}). 
A tidal interaction with its neighbor IC 5283 ($\sim$ 1.3$'$ away from NGC 7469) is believed 
to have caused the powerful SB activity (e.g., \citealt{1994AJ....108...90M,1995ApJ...444..129G}). 
Dusty, young ($<$ 100 Myr), and massive star clusters 
(individual typical mass is 10$^{6-7}$ $M_\odot$) exist within the ring (e.g., \citealt{2007ApJ...661..149D}). 

As for the central $\sim$ 1$''$ region, multi-wavelength studies including 
centimeter, K-band (2.2 $\mu$m), and 3.3 $\mu$m PAH feature indicate 
that the AGN is energetically dominant at these wavelengths 
(\citealt{2003ApJ...592..804L}; \citealt{1995ApJ...444..129G}; \citealt{2004ApJ...617..214I}), 
although one third of the K-band continuum is of stellar origin (\citealt{1995ApJ...444..129G}). 
A core jet-like structure (e.g., \citealt{2003ApJ...592..804L,2006ApJ...638..938A}) and ionized outflows 
(\citealt{2005ApJ...634..193S,2007A&A...466..107B}) have also been observed in this AGN. 
Reverberation mapping revealed the mass of the central supermassive black hole ($M_{\rm BH}$) to be 
$\sim$ 1 $\times$ 10$^7$ $M_\odot$ (\citealt{2004ApJ...613..682P,2014ApJ...795..149P}). 
This $M_{\rm BH}$ is typical for QSOs at $z$ = 5 on average, which will eventually 
grow up to $M_{\rm BH}$ = 10$^9$ $M_\odot$ at $z$ = 0 (\citealt{2009ApJ...690...20S}). 
In addition, the Eddington ratio of this AGN ($\sim$ 0.3; \citealt{2004A&A...413..477P}) is also comparable to those of QSOs. 
Therefore the obtained spectrum of NGC 7469 could serve as a local template for high redshift QSOs as well. 
Other relevant properties of NGC 7469 are summarized in Table \ref{tbl1} with the information shown above. 

\begin{table*}
\begin{center}
\caption{Properties of NGC 7469 \label{tbl1}}
\begin{tabular}{ccc}
\tableline\tableline
Parameter & Value$^a$ & Reference$^b$ \\
\tableline
RC3 morphology & (R$'$)SAB(rs)a & (1) \\
Position of the nucleus &  & (1) \\
$\alpha_{\rm J2000.0}$ & 23$^{\rm h}$03$^{\rm m}$15.6$^{\rm s}$ &  \\
$\delta_{\rm J2000.0}$ & +08$^\circ$52$'$26.4$''$ &  \\
Position angle [${}^\circ$] & 128 & (2) \\
Inclination angle [${}^\circ$] & 45 & (2) \\
Recession Velocity [km s$^{-1}$] & 4925 ($z$=0.0164) & (3) \\
Adopted distance [Mpc] & 70.1 &  \\
Linear scale [pc arcsec$^{-1}$] & 330 &  \\
Nuclear activity & Seyfert 1 & (4) \\
$L_{\rm 2-10keV}$ [erg s$^{-1}$] & 1.5 $\times$ 10$^{43}$ & (5) \\
$L_{\rm IR}$ [$L_\odot$] & 2.5 $\times$ 10$^{11}$ & (6) \\
$<$SFR$>$ (CND) [$M_\odot$ yr$^{-1}$ kpc$^{-2}$] & 50--100 & (7) \\
Stellar age (CND) [Myr] & 110--190 & (7) \\
\tableline
\end{tabular}
\tablecomments{$^{(a)}$ The LSR velocity of the center of the galaxy is obtained from the CO(1-0) emission. We use this value to calculate the distance to NGC 7469 and the linear scale, 
by assuming $H_0$ = 71 km s$^{-1}$ Mpc$^{-1}$, $\Omega_{\rm M}$ = 0.27, and $\Omega_{\rm \Lambda}$ = 0.73 cosmology. 
The $<$SFR$>$ (CND) indicates the averaged star formation rate over the central $\sim$ 1$''$ region (i.e., circumnuclear disk = CND). 
The stellar age is also for the CND. 
$^{(b)}$ (1) NASA/IPAC Extragalactic Database (NED, \url{http://ned.ipac.caltech.edu}); (2) \citet{2004ApJ...602..148D}; (3) \citet{1990ApJ...354..158M}; 
(4) \citet{1993ApJ...414..552O}; (5) \citet{2014ApJ...783..106L}; (6) \citet{2003AJ....126.1607S}; (7) \citet{2007ApJ...671.1388D}.
}
\end{center}
\end{table*}

Observations of CO molecules revealed the existence of 
a large amount of cold molecular gas (e.g., \citealt{1988ApJ...324L..55S,1990ApJ...354..158M,2009A&A...493..525I}), 
as well as warm molecular gas (\citealt{2015ApJ...801...72R}) in the central region of NGC 7469. 
The detailed spatial structure and the dynamics of the cold molecular gas were first investigated 
by high resolution (0.7$''$) CO(2-1) observations (\citealt{2004ApJ...602..148D}). 
Their CO(2-1) map also showed a ring-like structure with a similar radius 
seen at other wavelengths, and a bright extended ($\sim$ 1$''$) nuclear region. 
Note that we hereafter call this kind of (100 pc scale) central molecular concentration 
a {\it{circumnuclear disk (CND)}} in general. 
\citet{2004ApJ...602..148D} also found a bar or a pair of spiral arms between the ring and the CND, 
which is only visible in the molecular emission, and not prominent at NIR wavelengths. 
This configuration is quite unusual (see Section \ref{sec5}). 
The total molecular gas mass inside the $r$ $<$ 2.5$''$ (825 pc) region 
inferred from their observations is 2.7 $\times$ 10$^9$ $M_\odot$. 
The spatial distributions of typical dense gas tracers such as HCN and HCO$^+$, on the other hand, 
have not been clear due to the lack of spatial resolution of previous studies. 

In this paper, we present the spatial distribution and line ratios of submillimeter dense gas tracers 
based on our high resolution (0.50$''$ $\times$ 0.40$''$, 1$''$ = 330 pc) ALMA observations of 
the central kpc region of NGC 7469 involving HCN(4-3), HCO$^+$(4-3), CS(7-6), and partially CO(3-2) lines, 
and the underlying continuum emission. 
Our new ALMA observations improved this situation; 
we clearly separated the CND from the SB ring, and then obtained a spectrum of each component. 
The spatial resolution of this study (0.5$''$) is comparable to that of the 
CO(2-1) observation (0.7$''$) by \citet{2004ApJ...602..148D}, 
so that we can directly compare the molecular gas distributions and kinematics. 
Note that, however, the spatial resolution of this study ($\sim$ 150 pc) would still be insufficient 
to exclusively probe the region where gas heating due to an AGN with NGC 7469-like luminosity dominates (see Appendix-\ref{app-A} for details). 

We describe in Section \ref{sec2} the details of our observations. 
An 860 $\mu$m continuum map and a nuclear spectral energy distribution are shown in Section \ref{sec3}. 
A full band 7 spectrum and some line ratios are presented in Section \ref{sec4}, 
although we will extensively compare the line ratios with those obtained in other galaxies 
and discuss the possible causes of the HCN-enhancement in our succeeding paper (Izumi et al. submitted). 
The spatial distributions of the submillimeter dense gas tracers are shown in Section \ref{sec5}. 
Finally, our main conclusions are summarized in Section \ref{sec6}. 
A kinematic study using our data is presented in \citet{2015ApJ...806L..34F}.

\section{Observation and data reduction}\label{sec2}
\subsection{ALMA band 7 data of NGC 7469}
NGC 7469 was observed with ALMA on 2013 November 3-4 
with 28 antennas in C32-2 configuration, as a Cycle 1 early science program (ID = 2012.1.00165.S, PI = T. Izumi). 
Baseline lengths range from 15.0 to 1284.3 m (corresponds to the $uv$ distance from 17.4 to 1493.4 k$\lambda$ at 860 $\mu$m). 
The observations were conducted in a single pointing with a 18$''$ field of view, 
which fully covered the nucleus and the SB ring (total $\sim$ 4$''$ in diameter). 
We set the phase tracking center to ($\alpha_{\rm J2000.0}$, $\delta_{\rm J2000.0}$) 
= (23$^{\rm h}$03$^{\rm m}$15.64$^{\rm s}$, +08$^\circ$52$'$25.80$''$). 
The receiver was tuned to cover the redshifted lines of HCN(4-3) (rest frequency $\nu_{\rm rest}$ = 354.505 GHz), 
HCO$^+$(4-3) ($\nu_{\rm rest}$ = 356.734 GHz) in the upper side band, 
and CS(7-6) ($\nu_{\rm rest}$ = 342.883 GHz), CO(3-2) ($\nu_{\rm rest}$ = 345.796 GHz) in the lower side band, 
both in the 2SB dual-polarization mode. 
Each spectral window has a bandwidth of 1.875 GHz, 
and two spectral windows were set to each sideband 
to achieve a total frequency coverage of $\sim$ 7.5 GHz. 
The velocity spacing was originally 0.43 km s$^{-1}$ (488 kHz) per channel, 
but 47 channels were binned to improve the signal-to-noise (S/N) ratio, 
which resulted in a final velocity resolution of $\sim$ 20 km s$^{-1}$. 
In the following, we express velocities in the optical convention. 
Note that we could only cover the CO(3-2) line with $V_{\rm LSR}$ $>$ 4900 km s$^{-1}$ 
due to the limited band width for ALMA Cycle 1 observations, 
the recession velocity (4925 km s$^{-1}$, defined by CO(1-0) observations; \citealt{1990ApJ...354..158M}), 
and the molecular line width of NGC 7469 (typically full width at half maximum = FWHM $\sim$ 200 km s$^{-1}$ at the nucleus). 
The bandpass, phase, and flux were calibrated with 
3C454.3, J2257+0743 (this object is 2$^\circ$ away from NGC 7469), and J2232+117, respectively. 
Weather conditions were good throughout the observations 
with system temperatures of 100--200 K (usually 100--150 K). 
The total on-source integration time was $\sim$ 55 min. 

The reduction and calibration of the data were done with 
CASA version 4.1 (\citealt{2007ASPC..376..127M}) in standard manners, 
and the data was delivered to us from the East Asia ALMA Regional Center. 
All images of the line and continuum emissions were reconstructed 
with the CASA task \verb|CLEAN| (gain = 0.1, threshold = 1.0 mJy, weighting = natural). 
These images were further analyzed with MIRIAD (\citealt{1995ASPC...77..433S}). 
The achieved synthesized beams were typically 0.50$''$ $\times$ 0.40$''$, 
corresponding to 165 pc $\times$ 132 pc ($\sim$ 150 pc 
for the geometric mean) at the assumed distance of NGC 7469. 
The rms noise in the line channel maps is 0.55--0.58 mJy beam$^{-1}$ 
at a velocity resolution of 20 km s$^{-1}$ after a primary beam correction, 
and measured at areas free of line emission. 
However, in channels including CO(3-2) emission, 
the rms increases to $\sim$ 2.3 mJy beam$^{-1}$, probably due to remaining sidelobes. 
For the continuum emission, the channels free of line emission were averaged, 
which resulted in an rms of 0.09 mJy beam$^{-1}$ 
centered at $\nu_{\rm rest}$ = 349.7 GHz (860 $\mu$m), 
after combining both the LSB and USB data. 
This continuum emission was subtracted in the $uv$-plane before making line maps. 
The conversion factors between Jy beam$^{-1}$ and K are 51.50, 51.60, 50.47, and 51.83 K (Jy beam$^{-1}$)$^{-1}$ 
for CS(7-6), CO(3-2), HCN(4-3), and HCO$^+$(4-3), respectively. 

We used the MIRIAD task \verb|MOMENT| to make the integrated intensity maps presented in Section \ref{sec5}, 
by computing the zeroth moment of each data cube without any clipping. 
The velocity range included for the integration was 
$\sim$ 4650--5250 km s$^{-1}$ for CS(7-6), HCN(4-3), and HCO$^+$(4-3), 
but the lower limit was 4900 km s$^{-1}$ for CO(3-2) due to our spectral setting restrictions. 
These velocity ranges can cover the whole spectral range obtained within our 18$''$ field of view, except for CO(3-2). 
Throughout this paper, the pixel scale of ALMA images is set to 0.1$''$, 
and the displayed errors indicate only statistical ones unless mentioned otherwise. 
We use line intensities corrected for the primary beam attenuation for quantitative discussions. 
But this correction is not critical since most of the emission is within $\sim$ 4$''$. 
The adopted systematic error of the absolute flux calibration was 10 \%\footnote{\url{https://almascience.eso.org/documents-and-tools/cycle-1/alma-proposers-guide}}. 

\subsection{Ancillary data}
We retrieved the {\it{Hubble Space Telescope (HST)}} ACS/HRC image (F550M) of NGC 7469 
from the Hubble Legacy Archive (HLA)\footnote{\url{http://hla.stsci.edu}}. 
This image was calibrated by the HST pipeline products in a standard manner. 
We used the image only to check the positional consistency 
or shift between peaks in optical (V-band) and our ALMA band 7 images as shown in Section \ref{sec3}.

\section{860 $\mu$m continuum emission}\label{sec3}
\begin{figure*}
\epsscale{1}
\plotone{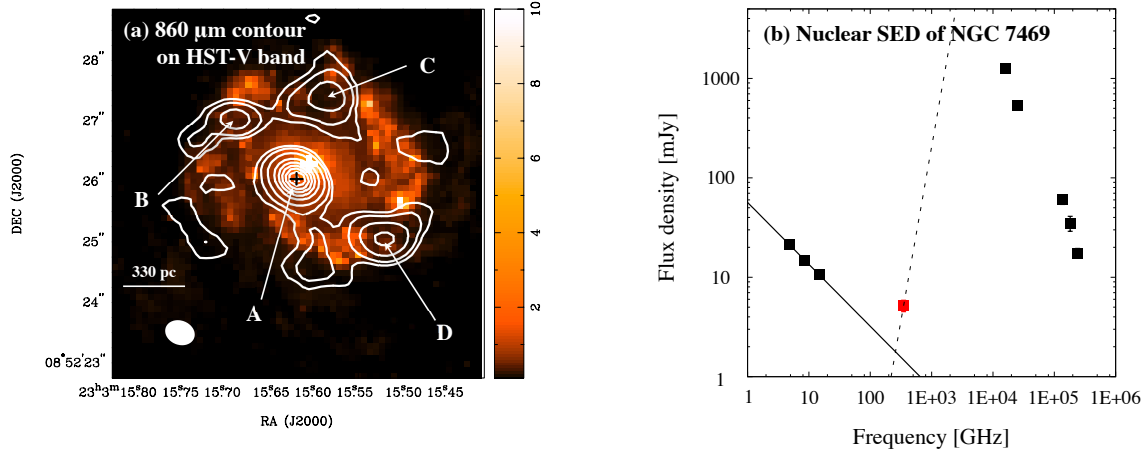}
\caption{(a) Continuum map towards the central 2 kpc region of NGC 7469 
at 860 $\mu$m (contours) overlaid on the HST-V band (F550M) map (color; the intensity scale only indicates the counts). 
Contours are 3, 5, 10, 20, 30, 40, and 50 $\sigma$, where 1 $\sigma$ = 0.09 mJy beam$^{-1}$. 
The maximum is 5.19 mJy beam$^{-1}$. 
The central cross indicates the peak position of the VLA 8.4 GHz continuum (\citealt{1991ApJ...378...65C}), 
which is identical to that of the 860 $\mu$m continuum. 
We regard this position as that of the AGN. 
The white filled ellipse indicates the synthesized beam at 860 $\mu$m (0.50$''$ $\times$ 0.40$''$, PA = 68.6$^\circ$). 
The AGN position and three bright knots in the SB ring are marked as A, B, C, and D, respectively.  
(b) The nuclear SED of the continuum emission in NGC 7469. 
We used 4.9 GHz data from \citet{1991ApJ...381...79W}, 8.4 GHz and 14.9 GHz data from \citet{2010MNRAS.401.2599O}, 
our ALMA band 7 data (red square), 18.7 $\mu$m and 11.8 $\mu$m data from \citet{2010MNRAS.402..879R}, and J, H, K-bands data from \citet{1995ApJ...444..129G}. 
The solid and dashed lines indicate $\nu^\alpha$ and $\nu^{2+\beta}$ scalings with $\alpha$ = -0.62 and $\beta$ = 1.5, respectively. 
The systematic flux errors are included in this plot. 
}
\label{figure1}
\end{figure*}

Figure \ref{figure1}a shows the 860 $\mu$m continuum emission 
overlaid on the HST V-band (F550M) emission in the central 2 kpc region of NGC 7469. 
In this figure, we can see a bright central source, 
which corresponds to the CND as defined in Section \ref{sec1.1}, 
and some knots in the SB ring with a radius of $\sim$ 1.5$''$ ($\sim$ 500 pc). 
The peak position of the 860 $\mu$m continuum emission is 
($\alpha_{\rm J2000.0}$, $\delta_{\rm J2000.0}$) 
= (23$^{\rm h}$ 03$^{\rm m}$ 15$^{\rm s}$.62, +08$^\circ$ 52$'$ 26.04$''$), 
which precisely coincides with that of the VLA 8.4 GHz (3.5 cm) 
continuum emission (\citealt{1991ApJ...378...65C,2010MNRAS.401.2599O}; 
their angular resolutions were 0.22$''$--0.25$''$). 
This precise match supports the accuracy of the astrometry of ALMA. 
We hereafter treat this peak position as the position of the AGN of NGC 7469. 
The 860 $\mu$m flux density there (marked as A in Figure \ref{figure1}a) 
is 5.19$\pm$0.09 mJy beam$^{-1}$. 
The 860 $\mu$m knots in the SB ring (marked as B, C, and D in Figure \ref{figure1}a) 
coincide with those visible at centimeter wavelengths 
(e.g., 8.4 GHz; \citealt{2010MNRAS.401.2599O}), 
and the knots of dense molecular gas tracers such as 
CO(3-2), HCN(4-3), and HCO$^+$(4-3) (Section \ref{sec5}). 
On the other hand, there is inconsistency between the peak positions 
of the 860 $\mu$m and those of the V-band emissions. 
The magnitude of this positional shift is $\sim$ 0.5$''$, 
which is the same size as our synthesized beam and thus significant. 
Similar discrepancy was also reported between the PdBI CO(2-1) emission 
and HST J-band maps (\citealt{2004ApJ...602..148D}), 
and between centimeter and J, H, K-bands maps (\citealt{1995ApJ...444..129G}), for example. 
Although one possible explanation for the discrepancy is 
that the absolute astrometry of HST is slightly off, 
it is rather likely that the stellar light is obscured by large amounts of dust 
traced by the 860 $\mu$m emission as also suggested by \citet{2007ApJ...661..149D}. 

We also investigated the origin (thermal or non-thermal) of the 860 $\mu$m continuum at the position-A 
by constructing the spectral energy distribution (SED) as shown in Figure \ref{figure1}b. 
We used high angular resolution (comparable to, or higher than our ALMA band 7 resolution = 0.5$''$) data of 
4.9 GHz from \citet{1991ApJ...381...79W}, 8.4 GHz and 14.9 GHz from \citet{2010MNRAS.401.2599O}, 
11.8 $\mu$m and 18.7 $\mu$m from \citet{2010MNRAS.402..879R}, and J, H, and K-bands from \citet{1995ApJ...444..129G}. 
The systematic flux uncertainty is included. 
Although the data is sparse at the submilimeter to FIR range, 
the SED indicates that the centimeter continuum emission can be 
well reproduced as synchrotron ($\alpha$ = -0.62$\pm$0.02; flux $\propto$ $\nu^\alpha$), 
whereas the flux density of the 860 $\mu$m continuum is well beyond the expected value of the synchrotron emission, 
i.e., the 860 $\mu$m continuum is dominated by thermal emission. 
Note that the percentage of the 860 $\mu$m continuum that is due to synchrotron is $\sim$ 30\% as estimated from the above scaling. 
In addition, we used the \verb|CLEAN| algorithm to deconvolve the continuum emission in the LSB and USB of ALMA band 7 separately. 
Then, the flux at position-A is 4.66$\pm$0.13 mJy beam$^{-1}$ and 5.00$\pm$0.13 mJy beam$^{-1}$ at 343.7 GHz and 356.0 GHz, respectively 
(errors indicate statistical ones only since they were observed simultaneously), suggesting a positive spectral index 
($\beta$ = 0$\pm$1; flux $\propto$ $\nu^{(2+\beta)}$, but the number of data points is just two for this fitting). 
Therefore, we conclude that the 860 $\mu$m continuum at the position-A mainly originates from thermal dust. 

\section{Band 7 spectra, line ratios, and dense molecular gas mass}\label{sec4}
\begin{figure*}
\epsscale{1}
\plotone{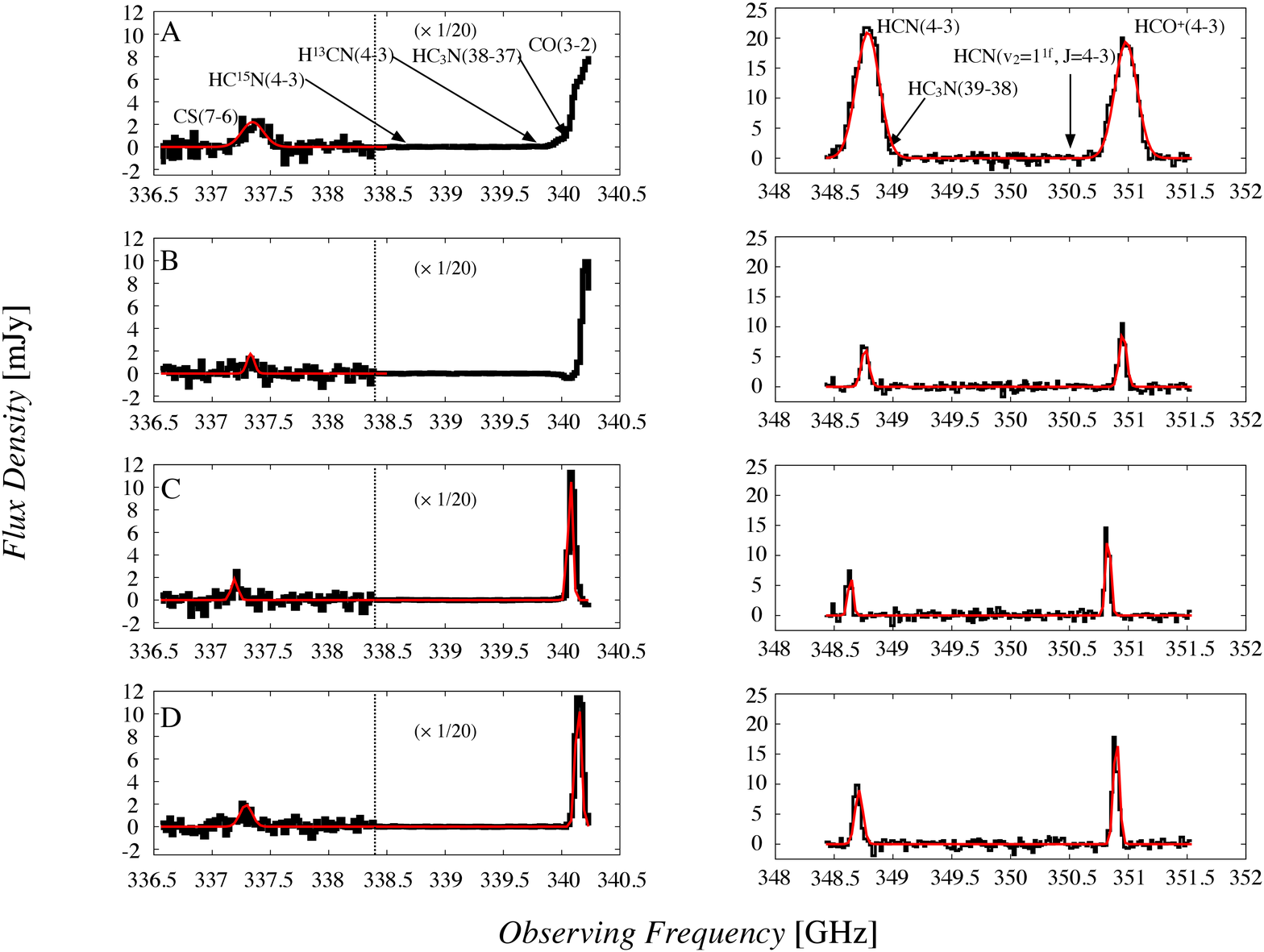}
\caption{ALMA band 7 spectra extracted with a single synthesized beam (150 pc-resolution) 
placed at positions-A, B, C, and D in Figure \ref{figure1}a. 
The flux of the spectral window containing CO(3-2) emission line is reduced by 20 times for a presentation. 
At the position-A (= AGN), we detected CO(3-2), HCN(4-3), HCO$^+$(4-3), and CS(7-6) with $>$ 3 $\sigma$ significance. 
Other lines were undetected (i.e., $<$ 3 $\sigma$) or blended with other lines. 
Their expected locations are marked by black arrows in the top panel. 
The redshift of the nucleus assumed in this study is $cz$ = 4925 km s$^{-1}$, 
based on CO(1-0) observations (\citealt{1990ApJ...354..158M}). 
The Gaussian fit to the observed CS(7-6), CO(3-2), HCN(4-3), and HCO$^+$(4-3) 
emissions is represented by the red line. 
The fitting parameters can be found in Table \ref{tbl2}. 
As for CO(3-2), we could not cover the whole spectral range in our observations. 
Thus, we use the peak flux density in the channel map (Figure \ref{figure3}) 
and the FWHM of HCN(4-3) emission line, to anyhow fit the CO(3-2) emission. 
However, we did not fit the CO(3-2) emission line at the positions A and B, 
since we could not find a clear turnover feature or blue part of the spectrum there. 
The continuum emission has been subtracted. 
Note that there is a dip at the lower frequency side of the CO(3-2) emission line 
at the position-B, which is due to a negative sidelobe (see also Figure \ref{figure3}). 
}
\label{figure2}
\end{figure*}

Here we show the ALMA band 7 spectra in Figure \ref{figure2}. 
These spectra were extracted from a single synthesized beam (150 pc-resolution) placed at the positions A-D in Figure \ref{figure1}a, for comparison. 
The strongest line at the position-A is CO(3-2), followed by HCN(4-3), HCO$^+$(4-3), and CS(7-6). 
Other lines were not detected (i.e., $<$ 3 $\sigma$) or are blended with other lines, 
and their spectral positions are marked by black arrows in Figure \ref{figure2}. 
On the other hand, at the SB ring (positions B-D), HCO$^+$(4-3) is brighter than HCN(4-3). 
The FWHM of the HCN(4-3) and HCO$^+$(4-3) emissions are 
200 and 190 km s$^{-1}$, 66 and 60 km s$^{-1}$, 41 and 42 km s$^{-1}$, and 50 and 50 km s$^{-1}$ at the positions A-D, 
respectively (see also Figures \ref{figure4} and \ref{figure5}). 
Thus, these values are consistent with each other, 
which suggests that they are tracing the same volume. 
We list the emission line parameters at these positions in Table \ref{tbl2} with other relevant quantities. 

\subsection{$R_{\rm HCN/HCO^+}$ and $R_{\rm HCN/CS}$}\label{sec4.1}
By inspecting the spectra, one may notice that the HCN(4-3)/HCO$^+$(4-3) integrated intensity ratio 
(hereafter $R_{\rm HCN/HCO^+}$) is higher than unity only at the position-A. 
To further clarify this, we list the values of $R_{\rm HCN/HCO^+}$, and the HCN(4-3)/CS(7-6) integrated intensity ratio 
(hereafter $R_{\rm HCN/CS}$; \citetalias{2013PASJ...65..100I} also proposed this ratio 
to be an empirical discriminator of AGN and SB) at the positions A-D in Table \ref{tbl3}. 
The integrated intensity of each line emission at each position is given in Table \ref{tbl2}. 
They are calculated by the Gaussian fit to the data. 
The line ratios in Table \ref{tbl3} are consistent to the submm-HCN enhancement 
scenario first proposed by \citetalias{2013PASJ...65..100I}, 
i.e., AGNs tend to show higher $R_{\rm HCN/HCO^+}$ 
and/or $R_{\rm HCN/CS}$ compared to SB galaxies, 
although the $R_{\rm HCN/HCO^+}$ at the position-A is only slightly larger than unity and totally comparable 
to the $R_{\rm HCN/HCO^+}$ of a typical early phase (e.g., \citealt{2009ApJ...694..610M}) 
SB galaxy NGC 253 (1.1$\pm$0.3; \citetalias{2013PASJ...65..100I}). 
But the ratio is significantly higher than that of a typical late phase (e.g., \citealt{2009ApJ...694..610M}) 
SB galaxy, M82 ($R_{\rm HCN/HCO^+}$ $\sim$ 0.4; \citetalias{2013PASJ...65..100I}). 
Note that a mechanical heating due to, e.g., frequent supernovae, 
would be responsible in the central region of NGC 253 (\citealt{2014A&A...564A.126R}) 
for its relatively high $R_{\rm HCN/HCO^+}$, as compared to M82. 

By investigating the 8.4 GHz radio flux obtained at 0.2$''$ (\citealt{2001ApJ...553L..19C}) and at 0.03$''$, 
\citet{2007ApJ...671.1388D} estimated that a nuclear supernova rate is $\sim$ 0.1 yr$^{-1}$. 
This rate corresponds to a star formation rate of $\sim$ 15 $M_\odot$ yr$^{-1}$ 
assuming the Salpeter initial mass function (\citealt{1955ApJ...121..161S}). 
Our relatively large spatial resolution ($\sim$ 150 pc) 
would pick up line fluxes emanating from the extended SB activity inside the CND 
(see also Appendix-\ref{app-A} for the estimated size of the XDR of NGC 7469), 
which would impact the line ratios discussed here. 
We can not find any morphological and kinematic signatures of a jet-ISM interaction 
in the high resolution maps of H$_2$ and Br$\gamma$ emissions (\citealt{2009ApJ...696..448H,2011ApJ...739...69M}), 
although \citet{2003ApJ...592..804L} reported the existence of a core radio jet-like structure. 
Therefore, we speculate that the energetics of the CND of NGC 7469 
is influenced by both the AGN (XDR) and the SB activity (PDR) 
when observed at the rather coarse resolution (150 pc in this work). 

In our succeeding paper (Izumi et al. submitted), 
we conducted non-local thermodynamic equilibrium (non-LTE) line modelings by the RADEX code (\citealt{2007A&A...468..627V}). 
Assuming a one zone cloud, we widely changed gas kinetic temperature, 
molecular column density, molecular abundance ratio, and background temperature 
(we expect higher background temperature in AGNs than in SB galaxies), 
but fixed H$_2$ gas density among the models ($n_{\rm H_2}$ = 10$^5$ cm$^{-3}$). 
This treatment was based on the assumption that 
there would be no systematic variation in $n_{\rm H_2}$ between AGNs and SB galaxies. 
Then, we calculated the resultant line intensity ratio, i.e., $R_{\rm HCN/HCO^+}$ or $R_{\rm HCN/CS}$, in each model case. 
As a result, we found that both $X$(HCN)/$X$(HCO$^+$) $\ga$ 10 
and $X$(HCN)/$X$(CS) $\ga$ 10 are required to reproduce 
$R_{\rm HCN/HCO^+}$ $>$ 1 and $R_{\rm HCN/CS}$ $>$ 10 as are observed in NGC 7469 and NGC 1097. 
Furthermore, the line opacity of HCN(4-3) is moderately thick, whereas that of CS(7-6) is thin when $R_{\rm HCN/CS}$ $\ga$ 10. 
Here, $X(\rm mol)$ indicates a molecular fractional abundance relative to H$_2$. 
The above required abundance ratios are significantly 
higher (at least by a factor of a few) than those necessary to reproduce the line ratios in SB galaxies. 
Therefore, these results suggest that a chemical abundance variation 
between AGNs and SB galaxies would be the key for the HCN-enhancement. 

\begin{table*}
\begin{center}
\caption{Parameters of detected lines at the peak positions of the 860 $\mu$m continuum emission \label{tbl2}}
\begin{tabular}{ccccccccc}
\tableline\tableline
Emission & $\nu_{\rm rest}$ & $n_{\rm crit}$ & $E_u/k_{\rm B}$ & Peak flux & $v_{\rm LSR}$ & $\Delta v$ & $S$ & $I$ \\
 & [GHz] & [cm$^{-3}$] & [K] & [mJy beam$^{-1}$] & [km s$^{-1}$] & [km s$^{-1}$] & [Jy beam$^{-1}$ km s$^{-1}$] & [K km s$^{-1}$] \\
 (1) & (2) & (3) & (4) & (5) & (6) & (7) & (8) & (9) \\
\tableline
CS($J$=7-6) & 342.883 & 2.9$\times$10$^6$ & 65.8 & 2.20$\pm$0.58 & 4919.7$\pm$29.3& 196.5$\pm$27.2 & 0.46$\pm$0.14 & 23.7$\pm$7.1 \\
 & & & & $<$1.74$^*$ & - & - & $<$0.12$^\dag$ & $<$6.2$^\dag$ \\
 & & & & 1.87$\pm$0.58 & 5059.9$\pm$60.6 & 59.0$\pm$18.0 & 0.12$\pm$0.05$^\ddag$ & 6.0$\pm$2.6$^\ddag$ \\
 & & & & 1.98$\pm$0.58 & 4982.2$\pm$47.9 & 74.1$\pm$17.0 & 0.16$\pm$0.06$^\ddag$ & 8.0$\pm$3.0$^\ddag$ \\ 
CO($J$=3-2) & 345.796 & 8.4$\times$10$^3$ & 33.2 & - & - & - & - & - \\
 & & & & 202$\pm$2$^{**}$ & - & - & {\textcolor{magenta}{-}} & {\textcolor{magenta}{-}} \\
 & & & & 232$\pm$2$^{**}$ & - & - & 9.7$\pm$1.2$^{\dag}$ & 498.9$\pm$61.0$^\dag$ \\
 & & & & 235$\pm$2$^{**}$ & - & - & 12.0$\pm$0.4$^{\dag}$ & 617.5$\pm$19.2$^\dag$ \\
HCN($J$=4-3) & 354.505 & 8.5$\times$10$^6$ & 42.5 & 21.00$\pm$0.55 & 4915.8$\pm$7.8 & 199.6$\pm$6.7 & 4.46$\pm$0.19 & 225.1$\pm$9.5 \\
 & & & & 6.41$\pm$0.55 & 4935.0$\pm$17.5 & 65.5$\pm$5.9 & 0.45$\pm$0.06 & 22.6$\pm$2.8 \\
 & & & & 6.84$\pm$0.55 & 5044.2$\pm$26.5 & 41.0$\pm$5.0 & 0.30$\pm$0.04 & 15.0$\pm$2.0 \\
 & & & & 8.98$\pm$0.55 & 4981.3$\pm$16.3 & 50.1$\pm$1.5 & 0.48$\pm$0.03 & 24.2$\pm$1.7 \\
HCO$^+$($J$=4-3) & 356.734 & 1.8$\times$10$^6$ & 42.8 & 19.38$\pm$0.58 & 4914.1$\pm$5.7 & 189.7$\pm$4.8 & 3.91$\pm$0.15 & 202.8$\pm$7.9 \\
 & & & & 9.09$\pm$0.58 & 4936.6$\pm$13.0 & 60.3$\pm$3.9 & 0.58$\pm$0.05 & 30.2$\pm$2.7 \\
 & & & & 14.50$\pm$0.58 & 5045.2$\pm$8.9 & 41.5$\pm$1.9 & 0.64$\pm$0.04 & 33.2$\pm$2.0 \\
 & & & & 18.24$\pm$0.58 & 4984.4$\pm$6.1 & 50.1$\pm$1.5 & 0.97$\pm$0.04 & 50.4$\pm$2.2 \\
860 $\mu$m & 349.7 & - & - & 5.19$\pm$0.09 & - & - & - & - \\
 & & & & 1.04$\pm$0.09 & - & - & - & - \\
 & & & & 1.23$\pm$0.09 & - & - & - & - \\
 & & & & 1.70$\pm$0.09 & - & - & - & - \\
\tableline
\end{tabular}
\tablecomments{Column 1: Full line name. Column 2: Line rest frequency. 
Column 3: Critical density of the line, calculated for $T_{\rm kin}$ = 100 K in the optically thin limit (\citealt{2009ApJ...692.1432G}). 
Column 4: Upper level energy. 
Column 5: Peak flux density of the line and continuum emission determined from the Gaussian fits to the data (Figure \ref{figure2}). 
They are extracted with a single synthesized beam (150 pc resolution) 
placed at the positions-A (AGN-position), B, C, and D (listed vertically in this order) marked in Figure \ref{figure1}a. 
For the lines, errors were estimated from the adjacent emission-free channels except for CO(3-2). 
For the continuum, an error was estimated from the emission-free areas in the continuum image. 
$^*$The 3 $\sigma$ upper limit is listed. 
$^{**}$The peak flux density (i.e., flux density at the turnover feature) in the channel map is listed. 
Thus, these are not the results of the Gaussian fit to the data. 
The primary beam attenuation is corrected. 
Column 6, 7: LSR velocity and full width at half maximum (FWHM) of the line, which is calculated by the Gaussian fit. 
Column 8, 9: Velocity-integrated intensity in the unit of [Jy beam$^{-1}$ km s$^{-1}$] and [K km s$^{-1}$] 
at the same positions as listed in Column 5, which is estimated by combining the Columns 5 and 7. 
$^\dag$The velocity centroid and the FWHM of HCN(4-3) 
at the corresponding position are used for the Gaussian fits. 
$^\ddag$Although the resultant integrated intensity is $<$ 3 $\sigma$, we list the value. 
}
\end{center}
\end{table*}

\begin{table*}
\begin{center}
\caption{HCN(4-3)/HCO$^+$(4-3), HCN(4-3)/CS(7-6), and HCN(4-3)/CO(3-2) integrated intensity ratios in NGC 7469 \label{tbl3}}
\begin{tabular}{ccccc}
\tableline\tableline
Position  & A & B & C & D \\
\tableline
$R_{\rm HCN/HCO^+}$ & 1.11$\pm$0.06 & 0.75$\pm$0.11 & 0.45$\pm$0.07 & 0.48$\pm$0.04 \\
$R_{\rm HCN/CS}$ & 9.50$\pm$2.87 & $>$3.65 & 2.50$\pm$1.13 & 3.03$\pm$1.15 \\
$R_{\rm HCN/CO}$ & - & -$^*$ & 0.03$\pm$0.01 & 0.04$\pm$0.01 \\
\tableline
\end{tabular}
\tablecomments{Row 1: The positions marked in Figure \ref{figure1}a, 
where line ratios are extracted with a single synthesized beam (150 pc-resolution). 
Row 2: HCN(4-3)/HCO$^+$(4-3) integrated intensity ratio. 
Row 3: HCN(4-3)/CS(7-6) integrated intensity ratio. 
Row 4: HCN(4-3)/CO(3-2) integrated intensity ratio (tentative). 
$^*$The peak intensity ratio at the position-B is $\sim$ 0.03. 
The integrated intensities of these lines are shown in Table \ref{tbl2}, 
although the values of CO(3-2) are tentative. 
The velocity ranges integrated over are shown in Section \ref{sec2}. 
The ratios are measured in the brightness temperature scale.}
\end{center}
\end{table*}

In the following, we discuss possible mechanisms to realize the abundance variation in AGNs, 
although it is still unclear, since (1) there are various ways to change the chemical compositions significantly 
(e.g., {\it{X-ray irradiation, mechanical heating, high temperature chemistry}}), 
and (2) the number of high resolution observational data of dense gas tracers is currently insufficient. 

In conventional XDR models (\citealt{1996A&A...306L..21L,2005A&A...436..397M}), 
we find reproducing a high $X$(HCN)/$X$(HCO$^+$) such as $\ga$ 10 is hard (usually it is less than unity), 
since a powerful ionizing capability in XDRs will favor to enhance ionic species such as HCO$^+$. 
Nonetheless, according to the steady state, gas phase 1-dimensional XDR models 
intensively presented by \citet{2005A&A...436..397M} and \citet{2007A&A...461..793M}, 
we can expect a high $X$(HCN)/$X$(HCO$^+$) and the resultant 
$R_{\rm HCN/HCO^+}$ $>$ 1 in a low-to-moderate gas density ($n_{\rm H_2}$ $\la$ 10$^5$ cm$^{-3}$) cloud 
with a low hydrogen column density ($N_{\rm H}$ $\la$ 10$^{22.5}$ cm$^{-2}$) 
and a high X-ray flux (e.g., $F_{\rm X}$ $\ga$ 10 erg s$^{-1}$ cm$^{-2}$). 
For larger $N_{\rm H}$ where the $F_{\rm X}$ is attenuated, they predicted $R_{\rm HCN/HCO^+}$ $<$ 1. 
Following this scenario, we estimate the largest distance from the AGN that can keep $F_{\rm X}$ $>$ 10 erg s$^{-1}$ cm$^{-2}$ 
is $r$ $\sim$ 18 pc and $\sim$ 35 pc for NGC 1097 and NGC 7469, respectively, 
considering their 2-10 keV X-ray luminosities 
($L_{\rm 2-10keV}$ = 6.9 $\times$ 10$^{40}$ erg s$^{-1}$ and 1.5 $\times$ 10$^{43}$ erg s$^{-1}$ 
for NGC 1097 and NGC 7469, respectively; \citealt{2014ApJ...783..106L}). 
We here assume that the X-ray emission is isotropic around the AGN itself. 
Attenuation due to obscuring material is considered for this estimation (\citealt{1983ApJ...270..119M}; Appendix-\ref{app-B}). 
But one critically weak point in this scenario is that the model predicted line intensities are 
extremely low (\citealt{2006ApJ...650L.103M,2007A&A...461..793M}), 
which is inconsistent to the prominent HCN(4-3) and HCO$^+$(4-3) emission lines in these galaxies. 
Furthermore, \citet{2015A&A...573A.116M} argued that HCN(1-0)/HCO$^+$(1-0) integrated intensity ratio 
is decreased in NGC 1097 as we approach the very central region 
($\sim$ 2 at the edge of the CND\footnote{An upper limit of the image convolved size is $\sim$ 90 pc $\times$ 70 pc (\citetalias{2013PASJ...65..100I})}, 
but $\sim$ 1.5 at the center), 
which seems to be incompatible with the above speculation of the largest distance for a high line ratio in NGC 1097. 
Note that, however, \citet{2015A&A...573A.116M} did not fully resolve the CND of NGC 1097, 
thus, their argument is a tentative one at this moment. 

Another possible scenario is the high temperature gas phase chemistry (\citealt{2010ApJ...721.1570H}), 
which significantly enhances $X$(HCN) especially at $T$ $>$ 300 K 
due to the activated formation path from CN (see also Figure 17 of \citetalias{2013PASJ...65..100I}). 
This mechanisms would be important to make line intensity of HCN(4-3) sufficiently detectable. 
Furthermore, $X$(HCO$^+$) will be somewhat decreased in these models since they will react with H$_2$O 
(this is also a typical molecule in a high temperature environment) to form H$_3$O$^+$. 
This type of chemistry can be complemented to XDRs (\citealt{2013ApJ...765..108H}) where we can expect 
much higher gas temperature in such regions due to efficient X-ray heating than in SB environments. 
Indeed, some fraction of warm molecular gas exists in NGC 7469, 
as evidenced by the detections of near-IR H$_2$ (\citealt{2009ApJ...696..448H}) 
and high-$J$ CO emission lines (\citealt{2015ApJ...801...72R}). 

With these items, we here focus on the notable fact that the $R_{\rm HCN/HCO^+}$ obtained at the position-A of NGC 7469 (1.11$\pm$0.13) 
is only as half as that at the CND of NGC 1097 (2.0$\pm$0.2; \citetalias{2013PASJ...65..100I}), 
despite the more than two orders of magnitude higher X-ray luminosity of NGC 7469. 
Note that NGC 1097 also hosts a nuclear SB activity in the CND ($\sim$ a few $M_\odot$ yr$^{-1}$; \citealt{2007ApJ...671.1388D}), 
which would contaminate the line fluxes obtained with the $\sim$ 100 pc resolution beam (\citetalias{2013PASJ...65..100I}). 
Furthermore, \citet{2014A&A...567A.125G} found $R_{\rm HCN/HCO^+}$ = 1.5 
at the vicinity of the AGN in NGC 1068 ($L_{\rm 2-10 keV}$ = 6.9 $\times$ 10$^{42}$ erg s$^{-1}$; \citealt{2006A&A...455..173P}), 
whereas it increases to $\sim$ 3 at the East/West-knots of its CND, which are $\sim$ 100 pc away from the AGN (their spatial resolution was $\sim$ 35 pc). 
These results indicate that higher X-ray luminosity does not simply lead to higher $R_{\rm HCN/HCO^+}$. 
From the perspective of high temperature chemistry, the results again seem to be controversial, 
since the large difference in X-ray luminosity would compensate the $\sim$ 2.5 times larger spatial scale (in area) 
sampled in NGC 7469 than in NGC 1097, 
leading to a higher X-ray energy deposition rate per particle 
(this is an important parameter to determine the thermal structure of XDRs e.g., \citealt{1996ApJ...466..561M}) in NGC 7469, 
if the molecular gas distributes uniformly over the CND. 
In this case, we can expect a higher $R_{\rm HCN/HCO^+}$ 
in NGC 7469 than in NGC 1097 by a naive view, but the actual result is totally opposite. 

Therefore, we presume that a non-radiative heating, namely mechanical heating 
caused by, e.g., AGN jet or outflow, is important to realize the high 
$R_{\rm HCN/HCO^+}$ of 2-3 (\citealt{2012A&A...542A..65K,2015A&A...574A.127K}) as is observed in 
NGC 1097 and in the East/West knots of the CND of NGC 1068. 
Indeed, a large radio jet exists in NGC 1068, which is supposed to cause shocks in the knots (e.g., \citealt{2011ApJ...736...37K}). 
Furthermore, a radio core jet has been discovered in NGC 1097 (e.g., \citealt{2014ApJ...787...62M}), 
although its intrinsic (deprojected) size is uncertain. 
However, we can not discard the possibility such as 
the line emitting region of NGC 1097 is extremely centrally-concentrated, 
which would result in higher energy deposition rate per particle (i.e., more violent X-ray heating) 
than in NGC 7469 and the close vicinity of NGC 1068. 
Moreover, we again emphasize that the line fluxes of NGC 7469 would be 
contaminated by the emission from the extended SB region inside the CND. 
Higher resolution observations of these molecular emissions are necessary to 
accurately map the distribution of the line ratios across the CND of NGC 1097 and NGC 7469, 
and to search for a signature of shock-ISM interaction. 

\subsection{Other line emissions and their ratios}\label{sec4.2}
Table \ref{tbl3} also includes HCN(4-3)/CO(3-2) integrated intensity ratios ($R_{\rm HCN/CO}$) extracted at the positions C and D. 
We exclude the ratio at the positions A and B since we could not measure the accurate fluxes there. 
Admitting the uncertainty in CO(3-2) integrated intensity (we used the velocity centroid and the FWHM of HCN(4-3) emission line 
to fit the Gaussian profile to CO(3-2) line; see also Figure \ref{figure2} and Table \ref{tbl2}), 
the tentative $R_{\rm HCN/CO}$ at the SB ring are 
well below the value obtained in Ori-KL ($\sim$ 0.32; \citealt{1997ApJS..108..301S}). 
Since the line critical density is orders of magnitude different between HCN(4-3) and CO(3-2) (see Table \ref{tbl2}), 
we can expect the line emitting region of HCN(4-3) is much smaller than that of CO(3-2). 
Hence, this ratio would be highly prone to be affected by the spatial resolution, 
which could lead to significantly lower $R_{\rm HCN/CO}$ in NGC 7469 than in Ori-KL. 
This is well manifested in the case of NGC 253 (SB galaxy). 
When observed at $\sim$ 200-300 pc resolution (comparable to our 150 pc resolution for NGC 7469), 
$R_{\rm HCN/CO}$ $\sim$ 0.03 (\citealt{2007ApJ...666..156K,1995A&A...302..343I}), which is quite similar to our result. 
However, the ratio increase to $\sim$ 0.08, i.e., increase by a factor of a few, when observed at higher resolution (25-30 pc; \citealt{2011ApJ...735...19S}). 
Then, matching the spatial resolution is important to compare line ratios of emission lines with totally different critical densities among objects.

We also found that the H${}^{12}$CN(4-3)/H${}^{13}$CN(4-3) isotopic ratio 
at the position-A was $\ga$ 11.9 in the brightness temperature scale. 
Note this is a peak intensity ratio and we used a 3 $\sigma$ upper limit for H${}^{13}$CN(4-3). 
Then, adopting a ${}^{12}$C/${}^{13}$C ratio of $>$ 40 observed in 
NGC 253 (\citealt{2014A&A...565A...3H}; \citealt{2010A&A...522A..62M}, $\sim$ 340 pc resolution), 
the upper limit to the opacity of the H$^{12}$CN(4-3) is 3.5 at the position.
For this calculation, we assume (1) the gas is in the LTE, (2) H$^{12}$CN is optically thick, 
(3) H$^{13}$CN is optically thin, and (4) H$^{12}$CN and H$^{13}$CN emissions are coming from the same volume of gas. 
Alternatively, if we adopt a ${}^{12}$C/${}^{13}$C ratio of 100 obtained in 
Mrk 231 (\citealt{2014A&A...565A...3H}, $\sim$ 19 kpc resolution), the upper limit of the opacity increases to 8.8. 
In either case, the derived optical depth is rather smaller than the values of the Galactic star forming regions 
(e.g., $\sim$ 70 in Ori-KL; \citealt{1997ApJS..108..301S}), 
likely reflecting the much more turbulent environments in the nuclear regions of galaxies. 

Note that vibrationally excited HCN emission, namely HCN($v_2$=1$^{1f}$, $J$=4-3) 
at $\nu_{\rm rest}$ = 356.256 GHz, was not detected at the position-A, 
despite the fact that NGC 7469 is classified as a LIRG (\citealt{2003AJ....126.1607S}) hosting a luminous AGN. 
Considering the above mentioned modest optical depth of H$^{12}$CN(4-3), 
we can expect HCN($v_2$=1$^{1f}$, $J$=4-3) is completely optically thin, 
which makes itself hard to be detectable. 
Therefore, the non-detection of HCN($v_2$=1$^{1f}$, $J$=4-3) emission does not necessarily mean 
that IR-pumping has little contribution to the observed line intensities, 
although we here simply treat that the HCN(4-3) emission 
is the consequence of the purely rotational transition. 
We need quantitative modeling of the line excitation to further address this problem. 
We also note that the contribution of HC$_3$N(39-38) emission to the HCN(4-3) emission is negligible at the position-A 
since HC$_3$N emission is typically much fainter than that of HCN, 
and the line profile of the bluer HCN(4-3) is very similar to that of HCO$^+$(4-3), which does not suffer from line blending. 
We thus conclude that the contribution of HC$_3$N(38-37) to the CO(3-2) emission is also negligible. 

\subsection{Dense molecular gas mass}\label{sec4.3}
In addition to $R_{\rm HCN/HCO^+}$ and $R_{\rm HCN/CS}$, 
we list line luminosities of HCN(4-3) and HCO$^+$(4-3) 
extracted at the position-A and the whole CND (i.e., central 1$''$ diameter region) 
in Table \ref{tbl4} as well since these lines are relatively strong enough to be detectable 
even in high redshift galaxies (e.g., \citealt{1997ApJ...484..695B,2011ApJ...726...50R}). 
By using this data, we roughly estimate the mass of dense molecular gas ($M_{\rm dense}$) in the CND as follows. 
First, we convert $L'_{\rm HCN(4-3)}$ to the corresponding value of HCN(1-0), 
assuming HCN(4-3)/HCN(1-0) = 0.6$\pm$0.2 in the brightness temperature unit. 
This ratio is an averaged value over the CND of the nearby typical Seyfert-2 galaxy, NGC 1068, 
sampled at 100 pc resolution (\citealt{2014A&A...570A..28V}). 
Note that NGC 1068 has a quite similar X-ray luminosity and an Eddington ratio to NGC 7469. 
Then, we obtain $L'_{\rm HCN(1-0)}$ = (9.3$\pm$3.1) $\times$ 10$^6$ K km s$^{-1}$ pc$^2$. 
Finally, assuming a conversion factor from $L'_{\rm HCN(1-0)}$ to $M_{\rm dense}$ 
to be 10, which is derived for AGNs by \citet{2008ApJ...677..262K} based on IRAM 30m observations, 
we estimate $M_{\rm dense}$ = (9.3$\pm$3.1) $\times$ 10$^7$ $M_\odot$. 
The dynamical mass ($M_{\rm dyn}$) of the CND, on the other hand, is calculated 
assuming a thin disk with Keplerian rotation as,  
$M_{\rm dyn}$ = 230 ($r$/pc) [$v(r)^2$/(km s$^{-1}$)] (sin $i$)$^{-2}$ $M_\odot$. 
Here, $v(r)$ is a rotation velocity at a radius $r$, and $i$ is an inclination angle of the disk. 
Then, putting the FWHM $\sim$ 200 km s$^{-1}$ at the CND (1$''$ = 330 pc diameter) 
and $i$ = 45$^\circ$ (\citealt{2004ApJ...602..148D}), 
we obtain $M_{\rm dyn}$ = 7.6 $\times$ 10$^8$ $M_\odot$. 
This is almost identical to the value derived in \citet{2015ApJ...806L..34F}, 
who isolated the rotation velocity and the velocity dispersion 
based on their harmonic decomposition analysis of the observed velocity field. 
Our $M_{\rm dyn}$ is also consistent to those derived for $r$ $<$ 2.5$''$ region, 
i.e., including the SB ring (6.5 $\times$ 10$^9$ $M_\odot$; \citealt{2004ApJ...602..148D}). 

Taking them together, we can estimate the dense gas mass fraction in the CND of NGC 7469 
as $f_{\rm gas}$ $\equiv$ $M_{\rm dense}/M_{\rm dyn}$ $\sim$ 12\%. 
Recalling the fact that the above used mass conversion factor 
has {\it{a factor of $\sim$ 3 uncertainty}} (\citealt{2008ApJ...677..262K}), 
$f_{\rm gas}$ eventually has the same level of uncertainty at this moment. 
Therefore, although the value of $f_{\rm gas}$ $\sim$ 12\% 
is comparable to those estimated for {\it{low-luminosity}} AGNs (\citealt{2012MNRAS.424.1963S}) using the same conversion factor (i.e., 10), 
it is still difficult to speculate whether or not $f_{\rm gas}$ has any relevance with the current mass accretion 
onto the central supermassive black hole. 
We need an accurate mass conversion factor estimated through multi-transitional analysis 
(i.e., estimating an accurate relative abundance of HCN with respect to H$_2$) to further 
constrain the $f_{\rm gas}$ and its relevance to the nuclear activity. 

\begin{table*}
\begin{center}
\caption{HCN(4-3) and HCO$^+$(4-3) line luminosities both in [$L_\odot$] unit and [K km s$^{-1}$ pc$^2$] unit \label{tbl4}}
\begin{tabular}{ccccc}
\tableline\tableline
 & $L_{\rm HCN(4-3)}$ & $L_{\rm HCO^+(4-3)}$ & $L'_{\rm HCN(4-3)}$ & $L'_{\rm HCO^+(4-3)}$ \\
 & [$L_\odot$] & [$L_\odot$] & [K km s$^{-1}$ pc$^2$] & [K km s$^{-1}$ pc$^2$] \\
\tableline
  AGN position & (7.95$\pm$0.34) $\times$ 10$^3$ & (7.01$\pm$0.27) $\times$ 10$^3$ & (5.58$\pm$0.24) $\times$ 10$^6$ & (4.83$\pm$0.19) $\times$ 10$^6$ \\
  CND & (2.10$\pm$0.09) $\times$ 10$^4$ & (1.86$\pm$0.08) $\times$ 10$^4$ & (1.47$\pm$0.06) $\times$ 10$^7$ & (1.28$\pm$0.06) $\times$ 10$^7$ \\
\tableline
\end{tabular}
\tablecomments{Row 1: Line luminosities within a synthesized beam (150 pc-resolution) placed at the position-A (i.e., AGN position) marked in Figure \ref{figure1}a. 
The luminosity was calculated following the formulae by \citet{2005ARA&A..43..677S}. 
Row 2: Line luminosities within the CND (i.e., the central $\sim$ 1$''$ = 330 pc diameter region).}
\end{center}
\end{table*}

\section{Spatial distribution of the dense gas tracers}\label{sec5}
\begin{figure*}
\epsscale{1}
\plotone{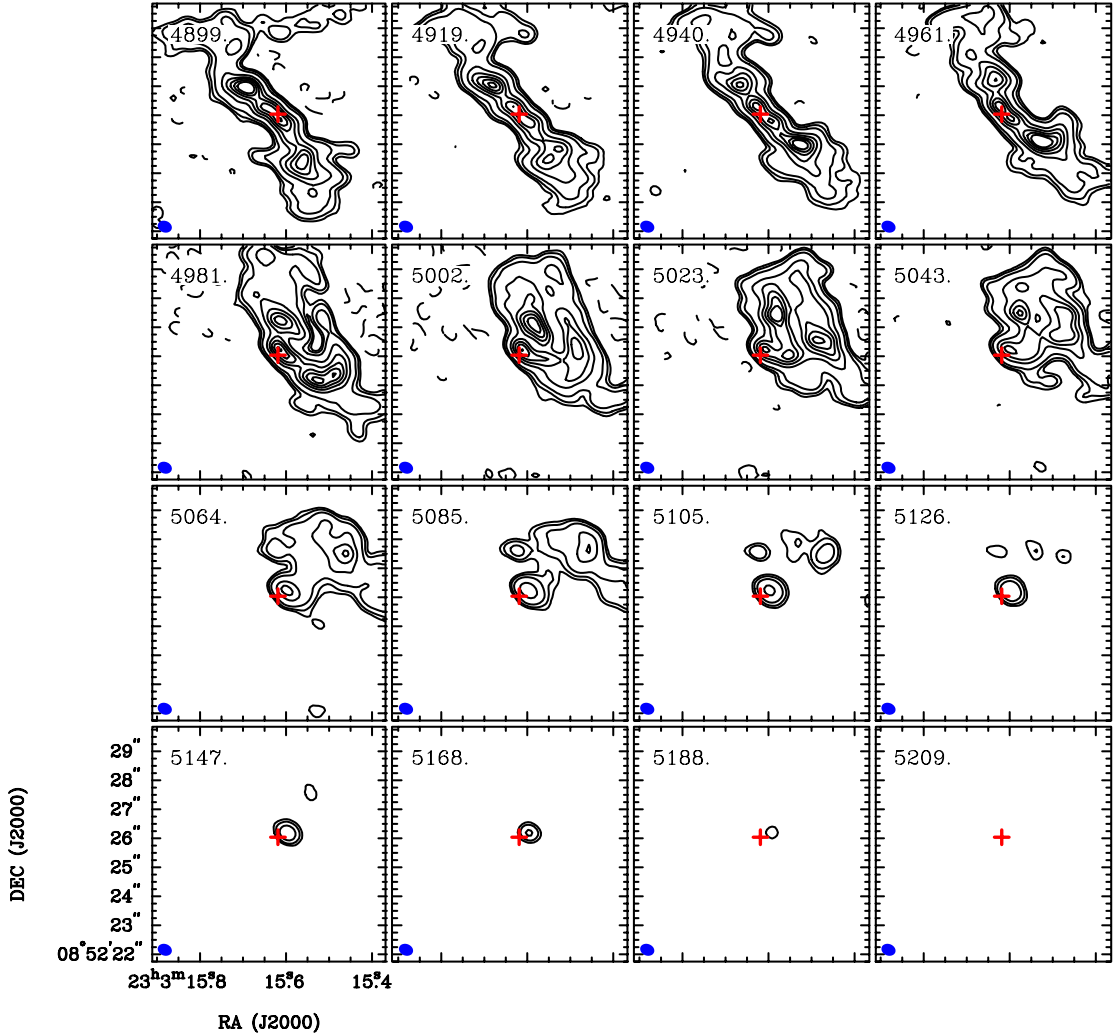}
\caption{Velocity channel maps of the CO(3-2) line emission in the central 8$''$ $\times$ 8$''$ (2.6 kpc $\times$ 2.6 kpc) region of NGC 7469. 
The central red cross in each channel indicates the position-A in Figure \ref{figure1}a (i.e., the AGN position). 
The velocity width of each channel is $\sim$ 20 km s$^{-1}$, and the central velocity of each channel is shown in the upper left corner. 
The synthesized beam (0.50$''$ $\times$ 0.40$''$, PA = 68.5$^\circ$) is also plotted in the bottom left corner. 
Contour levels are -5, 3, 5, 10, 30, 50, 70, 90, and 110 $\sigma$, where 1 $\sigma$ = 2.34 mJy beam$^{-1}$. 
The negative contour is indicated by the dashed lines. 
Note that we could not observe channels with $V_{\rm LSR}$ $<$ 4900 km s$^{-1}$ due to our observational setup. 
Attenuation due to the primary beam pattern is corrected. 
}
\label{figure3}
\end{figure*}

\begin{figure*}
\epsscale{1}
\plotone{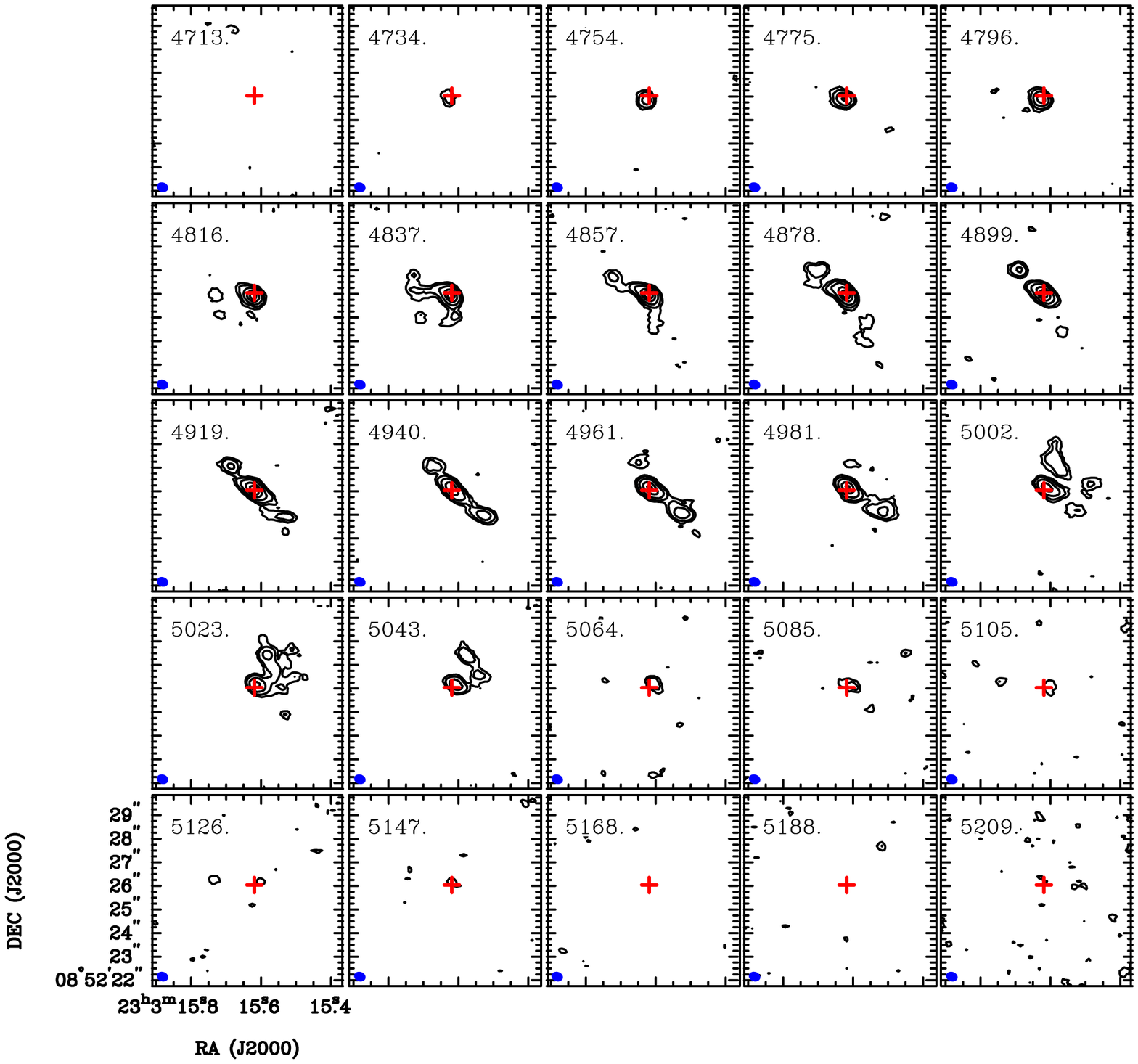}
\caption{Velocity channel maps of the HCN(4-3) line emission in the central 8$''$ $\times$ 8$''$ (2.6 kpc $\times$ 2.6 kpc) region of NGC 7469. 
The central red cross in each channel indicates the position-A in Figure \ref{figure1}a (i.e., the AGN position). 
The velocity width of each channel is $\sim$ 20 km s$^{-1}$, and the central velocity of each channel is shown in the upper left corner. 
The synthesized beam (0.49$''$ $\times$ 0.39$''$, PA = 70.4$^\circ$) is also plotted in the bottom left corner. 
Contour levels are -5, 3, 5, 10, 20, and 30 $\sigma$, where 1 $\sigma$ = 0.55 mJy beam$^{-1}$. 
Although we set the negative contours to be indicated by the dashed lines, they are not prominent in this figure. 
Attenuation due to the primary beam pattern is corrected. }
\label{figure4}
\end{figure*}

\begin{figure*}
\epsscale{1}
\plotone{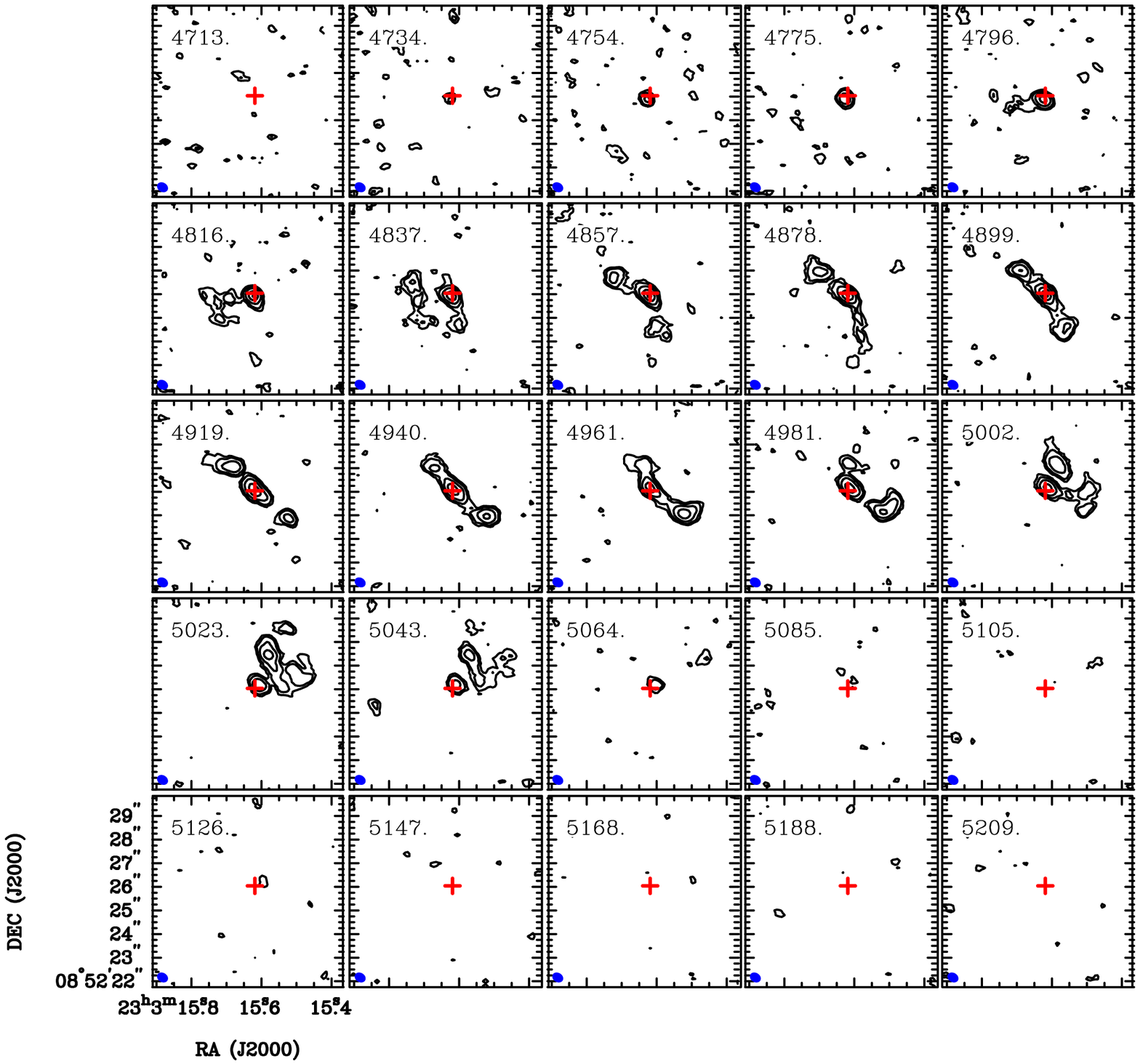}
\caption{Velocity channel maps of the HCO$^+$(4-3) line emission in the central 8$''$ $\times$ 8$''$ (2.6 kpc $\times$ 2.6 kpc) region of NGC 7469. 
The central red cross in each channel indicates the position-A in Figure \ref{figure1}a (i.e., the AGN position). 
The velocity width of each channel is $\sim$ 20 km s$^{-1}$, and the central velocity of each channel is shown in the upper left corner. 
The synthesized beam (0.48$''$ $\times$ 0.38$''$, PA = 57.9$^\circ$) is also plotted in the bottom left corner. 
Contour levels are -5, 3, 5, 10, 20, and 30 $\sigma$, where 1 $\sigma$ = 0.58 mJy beam$^{-1}$. 
Although we set the negative contours to be indicated by the dashed lines, they are not prominent in this figure. 
Attenuation due to the primary beam pattern is corrected. }
\label{figure5}
\end{figure*}

We show in Figures \ref{figure3} to \ref{figure5} the channel maps of the CO(3-2), HCN(4-3), and HCO$^+$(4-3) line emissions 
from the central 8$''$ $\times$ 8$''$ (2.6 kpc $\times$ 2.6 kpc) region of NGC 7469. 
We could not cover the whole CO(3-2) emission line as noted before. 
Comparing with CO(2-1) map in \citet{2004ApJ...602..148D}, 
we conclude that the emission from the south-west region of the map is missing. 
The HCN(4-3) and HCO$^+$(4-3) emissions were detected (i.e., $>$ 3 $\sigma$) over velocity ranges of 
4744-5165 km s$^{-1}$ and 4752-5150 km s$^{-1}$ at the position-A, respectively, 
of which widths (FWZI $\sim$ 400 km s$^{-1}$) are almost consistent with each other. 
As for CO(3-2), the covered velocity range corresponds to $\sim$ 70\% 
of the total velocity range of HCN(4-3) and HCO$^+$(4-3) at the position-A. 
However, we can not see a clear turnover feature in the CO(3-2) spectrum at that position (Figure \ref{figure2}). 
Therefore, we conclude that the FWZI of CO(3-2) emission line is significantly wider than those of HCN(4-3) and HCO$^+$(4-3) lines there. 
It might be possible that we detected CO(3-2) emission from an outer diffuse, high-velocity component around the CND. 
It might also be possible that the high velocity component of HCN(4-3) and HCO$^+$(4-3) emission lines 
are too faint to be significantly detectable. 
On the other hand, the line widths at the knots of the SB ring are almost consistent 
among CO(3-2), HCN(4-3), HCO$^+(4-3)$ emission lines (Figure \ref{figure2}). 
The channel maps also show that the CND and the SB ring are bridged by a bar-like, 
or several spiral arm-like structures as morphologically discussed later. 
A detailed kinematic study of this region is presented in \citet{2015ApJ...806L..34F}. 

\begin{figure*}
\epsscale{1}
\plotone{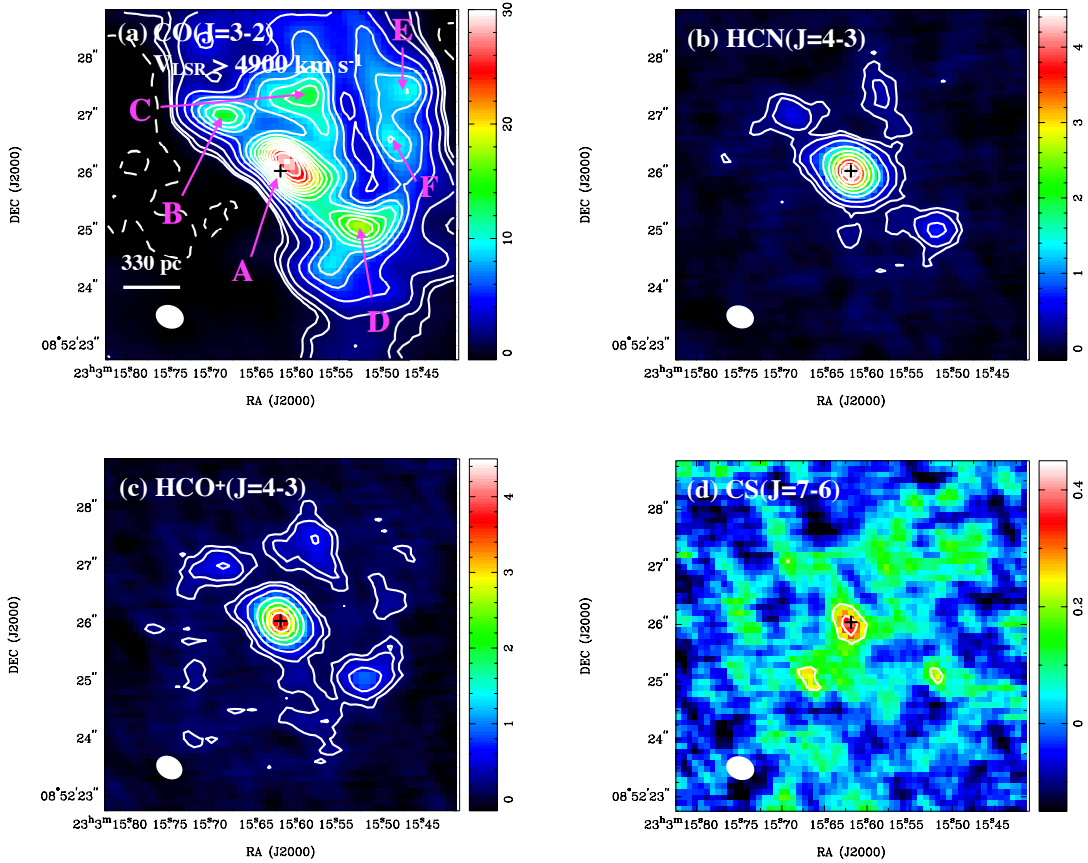}
\caption{Integrated intensity maps of (a) CO(3-2), (b) HCN(4-3), (c) HCO$^+$(4-3), and (d) CS(7-6), 
in the central $\sim$ 2 kpc region of NGC 7469. 
The rms noises (1 $\sigma$) are 0.20, 0.06, 0.06, and 0.06 Jy beam$^{-1}$ km s$^{-1}$ in (a), (b), (c), and (d), respectively. 
Note that we could not observe CO(3-2) emission at $V_{\rm LSR}$ $<$ 4900 km s$^{-1}$ 
due to our observational setting, thus the displayed map is incomplete (south-east part is deficient). 
Some bright molecular knots are marked as A-F in the CO(3-2) map (the positions A-D are the same as the ones marked in Figure \ref{figure1}a). 
The white filled ellipses indicate the synthesized beams ($\sim$ 0.50$''$ $\times$ 0.40$''$). 
The central crosses indicate the AGN position. 
Contours are: (a) -5, 3, 5, 10, 20, 30, ..., and 130 $\sigma$, (b) -5, 3, 5, 10, 20, 30, ..., and 70 $\sigma$, (c) -5, 3, 5, 10, 20, 30, 40, and 50 $\sigma$, (d) -5, 3, and 5 $\sigma$, respectively. 
Negative contours are indicated by dashed lines. 
See also Section \ref{sec2} for the velocity ranges integrated over. 
}
\label{figure6}
\end{figure*}

We also show the maps of the CO(3-2), HCN(4-3), HCO$^+$(4-3), and CS(7-6) velocity-integrated intensities 
in the central 2 kpc region of NGC 7469 in Figure \ref{figure6}. 
The peak positions of the 860 $\mu$m continuum emission (the positions A-D in Figure \ref{figure1}a) 
are again marked by the arrows in the CO(3-2) map. 
The dense gas tracers are also bright at these peak positions, 
indicating they are the sites of dust-obscured star formation. 
Two rather fainter knots are also seen in the CO(3-2) map, which are located northwest to the CND (marked as E and F). 

From these maps, it is apparent that the HCN(4-3), HCO$^+$(4-3), and CS(7-6) emissions are 
significantly concentrated towards the CND. 
A two-dimensional Gaussian fit to the HCN(4-3) image shows that the CND has an extent of 
(244.2$\pm$9.9) $\times$ (194.7$\pm$6.6) pc$^2$ with PA = 58.8$^\circ$$\pm$6.7$^\circ$, 
which is similar to that of HCO$^+$(4-3) emission. 
This is the beam-convolved size, thus the actual CND must be smaller. 
We derived these uncertainties using Monte Carlo techniques by creating 10$^3$ realizations of the image, 
each time adjusting the data by adding the normally-distributed (measured) noise randomly. 
Again using Figure \ref{figure6}, we estimate that the CND of NGC 7469 contains $\sim$ 82 \% and 53 \% 
of the total emission of HCN(4-3) and HCO$^+$(4-3) within the 18$''$ field of view of ALMA ($\sim$ 6 kpc), respectively, 
suggesting their utility to probe the nuclear regions of external galaxies especially about HCN(4-3). 

Here, we expect that the missing flux is not significant for the CND, 
although we recovered only 54$\pm$16 \% of the HCO$^+$(4-3) flux 
measured by APEX (\citealt{2014ApJ...784L..31Z}) within the central 18$''$ region, 
since its size ($\sim$ 1$''$) is much smaller than that of the maximum measurable 
angular scale of our observations, which was 7.1$''$ (2.3 kpc). 
Since the APEX's HCO$^+$(4-3) spectrum of 
\citet{2014ApJ...784L..31Z} is somewhat noisy, we can expect 
the systematic flux uncertainty would be larger 
than the value mentioned in that paper (15\% for absolute flux calibration uncertainty) 
additionally due to, e.g., baseline uncertainty. 
In other words, we expect the actual recovered rate can be higher than 54\%. 
Unfortunately, we can not estimate the percentage 
of the recovered flux of HCN(4-3) and CS(7-6) since there are currently 
no significant detection of these line emissions with single dish telescopes. 

In addition to the CND, we can see a partial circumnuclear SB ring at a radius of 1.5$''$--2.5$''$. 
\citet{2004ApJ...602..148D} also reported the existence of a bar or a pair of short, 
loosely wound spiral arms at a PA = 56$^\circ$ between the CND and the SB ring 
(connecting the positions B-A-D in Figure \ref{figure6}a), 
based on their 0.7$''$ CO(2-1) observations. 
Interestingly, that structure is only found in molecular emission; 
neither K-band (\citealt{1995ApJ...444..129G}; \citealt{2000AJ....119..991S}) 
nor centimeter continuum (\citealt{1991ApJ...378...65C}; \citealt{2010MNRAS.401.2599O}) image shows such a structure. 
This configuration is quite unusual if the structure is a genuine molecular bar, 
since a corresponding stellar structure should be easily visible in this scenario. 
Indeed, \citet{2004ApJ...602..148D} found no kinematic characteristics expected for a barred potential, 
i.e., an S-shaped contours in a isovelocity field (e.g., \citealt{2004A&A...422..865L,1999ApJ...511..157K}) 
and a tilted-X shape in a position-velocity diagram along the 
(hypothesized) bar axis (e.g., \citealt{1999ApJ...511..709L,2000ApJ...533..149S}). 
Moreover, Figures \ref{figure4} and \ref{figure5} reveal that there seem to be several connections 
between the CND and the SB ring in addition to the above hypothesized bar axis, 
e.g., channels at around 4850, 4980, and 5020 km s$^{-1}$. 
Although we can not rule out the existence of a faint stellar bar that could not be detected in previous NIR observations, 
it might be possible that the nuclear region of NGC 7469 consists of several spiral features 
like the case of, e.g., NGC 1097 (\citealt{2005AJ....130.1472P, 2006ApJ...641L..25F, 2009ApJ...702..114D}). 
Since the lack of spatial resolution leads to blending several molecular clouds, which would result in an illusional bar, 
higher angular resolution and sensitivity observations of both CO and NIR emissions are necessary to settle this issue. 

\section{Summary and conclusion}\label{sec6}
In this paper, we present high resolution (0.5$''$ $\times$ 0.4$''$; 1$''$ = 330 pc) ALMA band 7 (350 GHz band) 
observations of the submillimeter dense molecular gas tracers such as HCN(4-3), HCO$^+$(4-3), 
CS(7-6), and partially CO(3-2), in addition to the underlying 860 $\mu$m continuum emission, 
towards the central kpc region of the luminous type-1 Seyfert galaxy, NGC 7469. 
The region consists of the CND (central $\sim$ 1$''$) 
and the surrounding SB ring with a radius of 1.5$''$-2.5$''$. 
We revealed the spatial distribution of these emissions. 
Thanks to the high spatial resolution, we can reliably measure the line fluxes at the CND and the SB ring separately. 
The obtained line ratios are directly compared with those in NGC 1097 (a low-luminosity type-1 Seyfert galaxy), 
to achieve insights of possible effects of AGN luminosity on the surrounding molecular material. 
The main results of our observations and conclusions are summarized as follows: 

\begin{itemize}
\item[-] The spatial distribution of the 860 $\mu$m continuum emission 
resembles well to those of centimeter and MIR continuum emissions, and dense molecular gas tracers. 
But their peak positions (including some knots in the SB ring) 
are shifted significantly from the optical ones, 
suggesting the existence of dust obscured star formation there. 
\item[-] The maximum position of the 860 $\mu$m continuum is identical to that of the VLA 8.4 GHz continuum, indicating it is the AGN position. 
The SED reveals that the 860 $\mu$m continuum there is dominated by a thermal dust emission. 
We detected CO(3-2), HCN(4-3), HCO$^+$(4-3), and CS(7-6) emissions at the same position. 
Other lines including the vibrationally excited HCN were not detected, although NGC 7469 is classified as a LIRG. 
\item[-]  The high $R_{\rm HCN/HCO^+}$ and $R_{\rm HCN/CS}$ are only found in the CND, 
which is consistent to the findings in \citetalias{2013PASJ...65..100I}, 
i.e., AGNs tend to show higher $R_{\rm HCN/HCO^+}$ ($\ga$ 1) 
and/or $R_{\rm HCN/CS}$ ($\ga$ 10) than those in SB galaxies (submm-HCN enhancement). 
\item[-] Although the actual physical mechanisms to enhance $R_{\rm HCN/HCO^+}$ in AGNs are still unclear, 
we found this ratio is significantly lower in NGC 7469 (1.11$\pm$0.06) than 
in NGC 1097 (2.0$\pm$0.2) despite the more than two orders of magnitude 
higher X-ray luminosity of NGC 7469 than that of NGC 1097. 
The close vicinity of the luminous AGN of NGC 1068 also shows a comparable ratio ($\sim$ 1.5) to NGC 7469. 
Some other heating mechanisms than X-ray (e.g., mechanical heating) would 
contribute significantly for shaping the chemical composition in NGC 1097. 
\item[-] The CO(3-2) emission, which is not fully observed though, is distributed over both the CND and the SB ring, 
whereas the HCN(4-3) and the HCO$^+$(4-3) emissions are significantly concentrated toward the CND 
($\sim$ 82\% and 53\% of the total flux inside the 18$''$ field of view of ALMA, for HCN and HCO$^+$, respectively), 
suggesting their utility to probe the nuclear regions of galaxies. 
\item[-] Between the CND and the SB ring, we found several spiral-like structures, 
which would be transporting gas from the SB ring to the CND. 
This might be a more preferable scenario than a bar. 
We should test this tentative view by further high resolution observations and kinematic analysis. 
\end{itemize}

From this work, we achieved supportive evidence for the submm-HCN enhancement 
in AGNs tentatively proposed by \citetalias{2013PASJ...65..100I}. 
We will further investigate this trend by increasing the number of sample galaxies, 
and extensively discuss the possible cause of the HCN-enhancement 
in our succeeding paper (Izumi et al. submitted).

\acknowledgments
We thank the anonymous referee for very kind and helpful comments for improving the paper. 
This paper makes use of the following ALMA data: ADS/JAO.ALMA\#2012.1.00165.S. 
ALMA is a partnership of ESO (representing its member states), NSF (USA), and NINS (Japan), 
together with NRC (Canada) and NSC and ASIAA (Taiwan), in cooperation with the Republic of Chile. 
The Joint ALMA Observatory is operated by ESO, AUI/NRAO, and NAOJ. 
The National Radio Astronomy Observatory is a facility of the National Science Foundation 
operated under cooperative agreement by Associated Universities, Inc. 
We used data based on observations with the NASA/ESA Hubble Space Telescope, 
and obtained from the Hubble Legacy Archive, 
which is a collaboration between the Space Telescope Science Institute (STScI/NASA), 
the Space Telescope European Coordinating Facility (ST-ECF/ESA) 
and the Canadian Astronomy Data Centre (CADC/NRC/CSA). 
In addition, this research has made use of the NASA/IPAC Extragalactic Database (NED) 
which is operated by the Jet Propulsion Laboratory, California Institute of Technology, 
under contract with the National Aeronautics and Space Administration. 
T. I. was supported by the ALMA Japan Research Grant of NAOJ Chile Observatory, NAOJ-ALMA-0029. 
T. I. and H. U. are thankful for the fellowship received from the Japan Society for the Promotion of Science (JSPS).

\bibliography{TakumaIzumi}

\appendix
\section{A. The size of an XDR}\label{app-A}
Here we estimate the size of an XDR, 
which is defined as the largest distance from an AGN where X-ray heating due to the AGN 
dominates over UV heating due to SB, by employing a toy model. 

\begin{figure*}
\epsscale{0.6}
\plotone{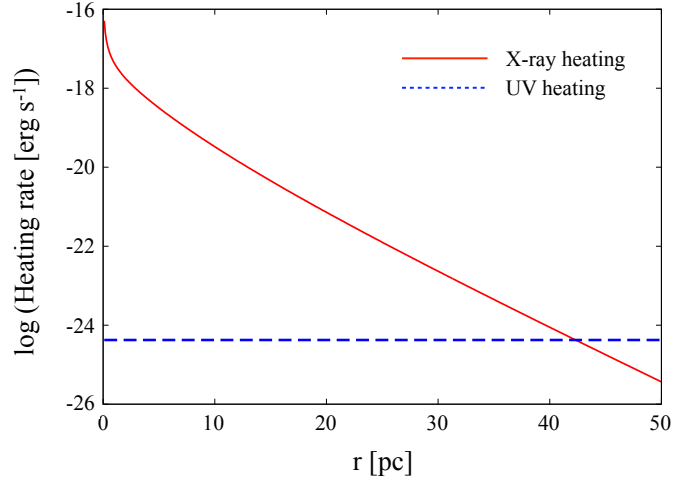}
\caption{
The heating rate per hydrogen atom due to X-ray absorption (red solid) 
and to soft-UV absorption (blue dashed) as a function of radius from the center, under the 1D slab geometry. 
We assume that a $M_{\rm BH}$ = 10$^7$ $M_\odot$ black hole is emitting 3\% of its Eddington luminosity at 1-10 keV. 
The soft-UV radiation field is taken as $G_0$ = 10$^{2.5}$ (Habing unit). 
The adopted gas density is $n_{\rm H}$ = 10$^5$ cm$^{-3}$. 
In this configuration, the estimated size of the XDR is $r$ $\sim$ 42 pc. 
}
\label{figure7}
\end{figure*}

For the X-ray incident flux, we adopt a spectral shape of the form, 
\begin{eqnarray}\label{eq.A1}
F(E) = F_0 \left(\frac{E}{\rm 1\hspace{0.5mm}keV}\right)^{- \alpha} \exp(-E/E_c), 
\label{eq_a1}
\end{eqnarray}
where $F_0$ is a scaling constant, $E$ = $h\nu$ eV, $\alpha$ is a power-law index, $E_c$ is a cut-off energy. 
We set $\alpha$ = 1.0 (e.g., \citealt{1995ApJ...438..672M,1997iagn.book.....P}) and $E_c$ = 100 keV (e.g., \citealt{1995ApJ...438..672M}). 
We hereafter consider the 1-10 keV photons for simplicity. 
The softer X-ray can be easily absorbed at the edge of the molecular cloud and thus neglected here. 
Considering the situation that an AGN with $r$ = 0.1 pc and $M_{\rm BH}$ = 10$^7$ $M_\odot$, 
is emitting isotropically at 3\% of its Eddington luminosity at this 1-10 keV range, we scaled $F_0$. 
Subsequent radiative transfer is calculated by adopting a 1D slab geometry with a uniform gas density of $n_{\rm H}$ = 10$^5$ cm$^{-3}$. 
Note that the AGN of NGC 7469 hosts a $\sim$ 10$^7$ $M_\odot$ black hole with the Eddington ratio of $\sim$ 0.3 (see also Section \ref{sec1}). 
Thus, our estimation corresponds to the case where 10\% of the bolometric AGN luminosity of NGC 7469 is going into the 1-10 keV band. 
Then, an energy deposition rate per particle ($H_{\rm X}$; \citealt{1996ApJ...466..561M}) is 
\begin{eqnarray}\label{eq.A2}
H_{\rm X} = \int_{\rm 1\hspace{0.5mm}keV}^{\rm 10 \hspace{0.5mm}keV} \sigma_{\rm pa}(E) F(E) \exp(-\tau) dE, 
\label{eq_a2}
\end{eqnarray}
with an optical depth of 
\begin{eqnarray}\label{eq.A1}
\tau = \sigma_{\rm pa}(E) \cdot N_{\rm H}(r). 
\label{eq_a3}
\end{eqnarray}
Here, we use the photoelectric absorption cross section $\sigma_{\rm pa}(E)$ of \citet{1983ApJ...270..119M}, 
which is estimated for 1-10 keV range. 
The attenuating hydrogen column density from the AGN to the point at the distance of $r$ pc is indicated by $N_{\rm H}$. 
Assuming the heating efficiency of X-ray photons to be 30\% (e.g., \citealt{1996ApJ...466..561M,2005A&A...436..397M}), 
an X-ray heating rate is calculated as a function of a radius from the AGN (Figure \ref{figure7}). 
On the other hand, we adopt a typical energy of 10 eV and a cross section of 2.78 $\times$ 10$^{-22}$ cm$^2$ 
for soft-UV photons from the SB activity (\citealt{2005A&A...436..397M}). 
This SB is assumed to produce a soft-UV radiation field of $G_0$ = 10$^{2.5}$ (Habing unit). 
Note that \citet{2014A&A...564A.126R,2014A&A...568A..90R} used this $G_0$ to reproduce $J_{\rm upper}$ $\la$ 4 $^{12}$CO 
emission lines (i.e., lines tracing cold molecular gas) observed in NGC 253 and Arp 299.  
Assuming that the UV heating efficiency is $\sim$ 0.3\%, 
the heating rate due to SB-induced UV radiation is also shown in Figure \ref{figure7}. 

Although the above estimation strongly depends on the adopted parameters and the value of $G_0$ 
is not well constrained in the circumnuclear region of Seyfert galaxies at this moment, 
we can roughly estimate that the size of the XDR from Figure \ref{figure7}, which is $r$ $\sim$ 42 pc in this case. 
This size is well smaller than the spatial resolution of our ALMA observations (150 pc).

\section{B. Attenuated X-ray flux}\label{app-B}
\begin{figure*}
\epsscale{0.6}
\plotone{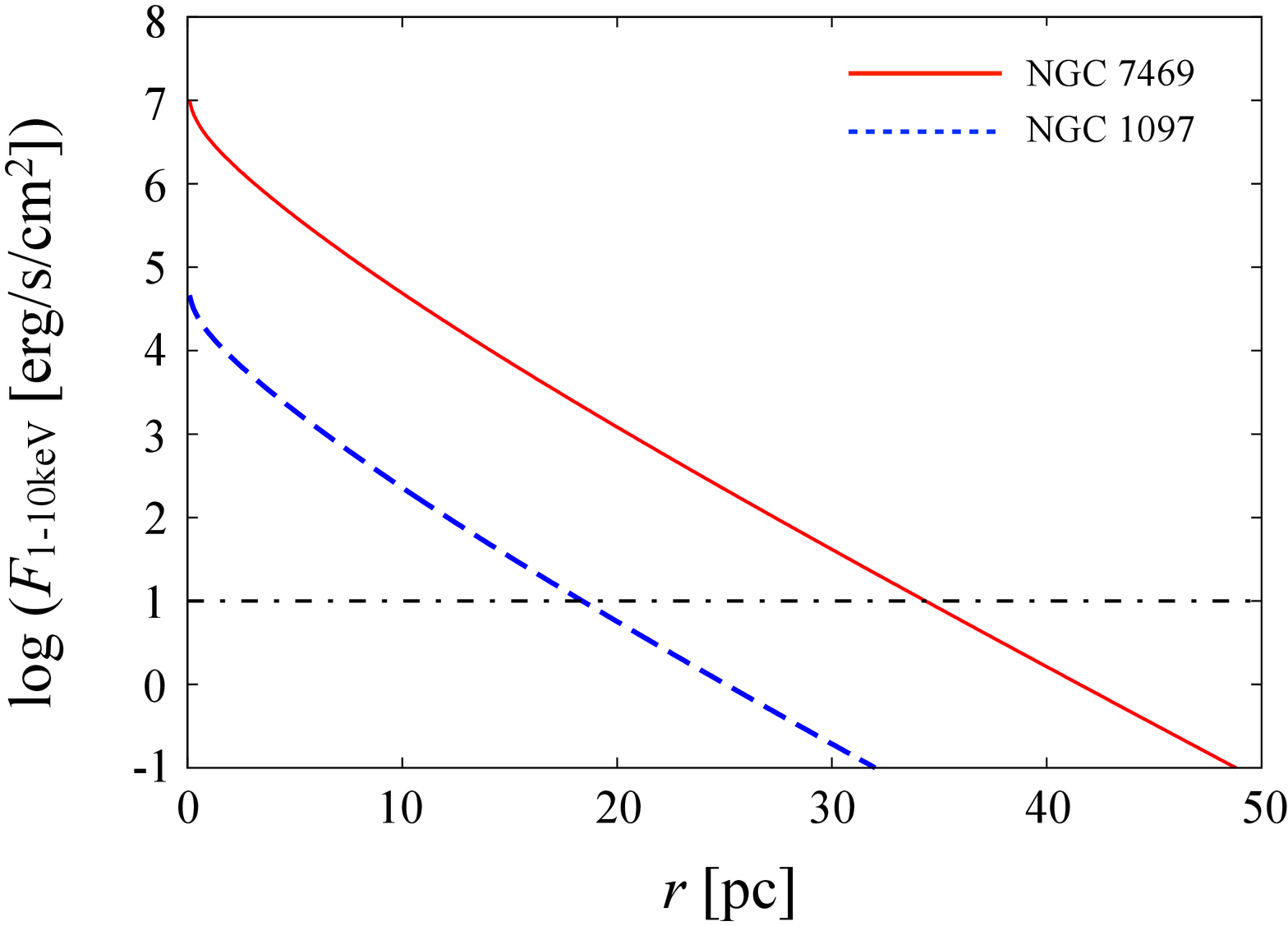}
\caption{
1D calculation of the 1-10 keV X-ray flux as a function of radius from the AGN. 
Attenuation due to obscuring material (\citealt{1983ApJ...270..119M}) is taken into account. 
The cases of NGC 7469 (red solid) and NGC 1097 (blue dashed) are shown. 
The horizontal dot-dashed line indicates the flux of 10 erg s$^{-1}$ cm$^{-2}$, 
which is a required value by \citet{2007A&A...461..793M} to realize $R_{\rm HCN/HCO^+}$ $>$ 1. 
Thus, the largest radius that can keep $F_{\rm 1-10 keV}$ $\ge$ 10 is $\sim$ 35 pc and $\sim$ 18 pc, 
for NGC 7469 and NGC 1097, respectively. 
}
\label{figure8}
\end{figure*}

Using the Equations (\ref{eq_a1}) and (\ref{eq_a3}), we can calculate an attenuated 1-10 keV X-ray flux as a function of radius. 
The cases of NGC 7469 and NGC 1097 are investigated. 
Again we used the photoelectric absorption cross section calculated by \citet{1983ApJ...270..119M}. 
We adopt the same values as indicated in Appendix-\ref{app-A} for this calculation, except for $F_0$. 
We this time scaled $F_0$ by using the 2-10 keV luminosity of NGC 7469 and NGC 1097 (e.g., \citealt{2014ApJ...783..106L}), 
$\alpha$ = 1.0, and assuming their AGN sizes to be $r$ = 0.1 pc. 
Then, the attenuated X-ray flux is shown in Figure \ref{figure8}. 

\end{document}